\documentclass[a4paper,11pt]{article}
\pdfoutput=1
\usepackage{jcappub}
\usepackage[T1]{fontenc}
\usepackage{mathtools,bm,booktabs,enumitem}

\newcommand{\mpl}{m_{\rm pl}}
\newcommand{\Mpl}{M_{\rm pl}}

\newcommand{\rhoMH}{\rho_M^H}
\newcommand{\nMH}{n_M^H}
\newcommand{\GaM}{\Gamma_M}
\newcommand{\Gde}{\Gamma_M^{\rm de}}
\newcommand{\Ocal}{\mathcal{O}}
\newcommand{\kk}{\mathbf{k}}
\newcommand{\xx}{\mathbf{x}}
\newcommand{\dd}{\mathrm{d}}

\newcommand{\hf}{h_f^{\rm fast}}
\newcommand{\Ups}{\Upsilon}
\newcommand{\Gtot}{\Gamma_{\rm tot}}

\title{\boldmath Cosmic rate equation and massive particle--antiparticle pair production}

\author{She-Sheng Xue}
\affiliation{ICRANet, Piazzale della Repubblica 10, 65122 Pescara, Italy
\\ Physics Department, Sapienza University of Rome, 
Rome, Italy\\INFN, Sezione di Perugia, 
Perugia, Italy
\\ICTP-AP, University of Chinese Academy of Sciences, Beijing, China
}

\emailAdd{shesheng.xue@icranet.org, shesheng.xue@gmail.com}

\abstract{
In a Friedmann universe with a time-varying cosmological term, the production of massive particle--antiparticle
pairs and their exchange of energy with the spacetime vacuum are described by a cosmic rate equation,
$\dot\rho_M+3H\rho_M=\Gamma_M(\rho_M^H-\rho_M)-\Gamma_M^{\rm de}\rho_M$. Together with the Friedmann equations, an energy-balance equation for the time-varying cosmological term, and a reheating equation for the radiation,
it forms a closed set of four independent background equations. We derive this set from a complex scalar field with a global U(1)
symmetry and a non-minimal coupling to curvature. In the adiabatic case, the occupation of non-relativistic modes
gives $\rho_M^H=2\chi m^2H^2$. A kinetic equation, based on the pairwise coupling $\xi R|\Phi|^2$ to fluctuations of the spacetime vacuum, gives the
relaxation rate $\Gamma_M$, and a Yukawa coupling gives the decay rate $\Gamma_M^{\rm de}$. In the fast-oscillating case, the Hubble function oscillates at
twice the pair mass and locks in phase with the pairs, so that it exchanges no net energy with them. The energy of the
created pairs is supplied instead by the decrease of the cosmological term in the time evolution,
$-\dot\rho_\Lambda=[(3-2\epsilon) H+\Gamma_M^{\rm de}]\rho_M^H$. Averaging over the fast time gives the cosmic rate equation in the
detailed-balance regime $\Gamma_M\gg H$. In this regime the closed set has an exact power-law solution with $\epsilon=-\dot H/H^2=\chi m^2/m_{\rm pl}^2$. We
discuss the application of the closed set to inflation, reheating and standard cosmology.
}


\begin{document}
\maketitle
\flushbottom

\section{Introduction}\label{sec:intro}

The creation of particles by the expansion of the Universe is one of the basic predictions of quantum
field theory in curved spacetime~\cite{Parker1968,Parker1969,ZeldovichStarobinsky1972,BirrellDavies,ParkerToms,ParkerFulling1973}.
A quantum field whose mode frequencies depend on time is not left in the spacetime vacuum: positive- and
negative-frequency solutions mix, and the Bogoliubov coefficients that describe the mixing count the pairs
created by the background. For a field of mass $m$ in de Sitter space with $m\gg H$ the number of created
pairs is suppressed as $e^{-2\pi m/H}$~\cite{GibbonsHawking1977,BunchDavies1978,Mottola1985}. Gravitational
particle production at the end of inflation~\cite{Ford1987} has been studied extensively as a source of
superheavy dark matter and other relics~\cite{ChungKolbRiotto1999,KuzminTkachev1998,ChungCrottyKolbRiotto2001,KolbLong2024}.
When the Hubble function oscillates, as it does during coherent oscillations of the inflaton, production
is enhanced~\cite{EmaJinnoMukaidaNakayama2015,EmaJinnoMukaidaNakayama2015b,EmaJinnoMukaidaNakayama2016,EmaNakayamaTang2018}.
These processes form part of the broader theory of inflation~\cite{Starobinsky1980,Guth1981,Linde1982,AlbrechtSteinhardt1982}
and of reheating~\cite{AbbottFarhiWise1982,AlbrechtSteinhardtTurnerWilczek1982,DolgovKirilova1990,TraschenBrandenberger1990,KofmanLindeStarobinsky1994,ShtanovTraschenBrandenberger1995,KofmanLindeStarobinsky1997,AllahverdiEtAl2010,AminEtAl2015,KolbTurner}.

A related line of work considers a cosmological term that varies in time and exchanges energy with matter.
Examples are decaying-vacuum cosmologies~\cite{OzerTaha1986,Lima1996,OverduinCooperstock1998,ShapiroSola2002}.
In the scenario of Refs.~\cite{Xue2015,Xue2023,Xue2023b,Xue2024}, the time-varying cosmological term drives inflation and,
as it decreases, produces massive particle--antiparticle pairs from the spacetime vacuum, which bring inflation to an end. During reheating, the pairs
oscillate back and forth between creation and annihilation, an exchange described by a cosmic rate
equation,
\begin{equation}
  \dot\rho_M+3H\rho_M=\GaM\big(\rhoMH-\rho_M\big)-\Gde\,\rho_M ,
  \label{eq:rate_intro}
\end{equation}
where $\rhoMH\simeq2\chi m^2H^2$ is the pair density produced by the spacetime vacuum, $\GaM$ the rate of the back-and-forth
exchange and $\Gde$ the decay rate into relativistic particles. Together with the Friedmann equations, an energy-balance equation for the time-varying cosmological term, and a reheating equation for the radiation, Eq.~\eqref{eq:rate_intro} closes the background equations, which
describe the evolution of the Universe.

In this article we derive this closed set of background equations from an explicit field theory: a complex
scalar field with a global U(1) symmetry and a non-minimal pairwise coupling $\xi R|\Phi|^2$ to the curvature, in a flat Friedmann universe
with a time-varying cosmological term. Our purpose is limited to the background: we derive the closed set of
four equations and examine its assumptions. We do not discuss perturbations, inhomogeneities, or the
numerical values of temperatures. The article is in two parts.

\emph{Part I (adiabatic case)}, sections~\ref{sec:frw}--\ref{sec:closed1}. Section~\ref{sec:frw} derives the Friedmann
equations from Einstein's equations with $\Lambda(t)$ and shows why a time-varying cosmological term requires an energy
exchange with matter. Section~\ref{sec:field} introduces the scalar field and shows that, after a rescaling, the Friedmann
background couples to it only through a pairwise term; it also derives the exact kinetic equations that govern the
occupation and the pair coherence of each mode. Section~\ref{sec:prod} derives the pair density $\rhoMH=2\chi m^2H^2$ from the
adiabatic occupation, the relaxation term of Eq.~\eqref{eq:rate_intro} from a kinetic equation for the coupling to
fluctuations of the spacetime vacuum, and the decay term from a U(1)-invariant Yukawa coupling. Section~\ref{sec:closed1} gives the resulting closed set of four equations and its exact solution in the
detailed-balance regime. It then discusses the application of the equations to inflation and to standard cosmology after
reheating, and gives three arguments for why the relaxation rate is proportional to the pair mass.

\emph{Part II (fast-oscillating case)}, sections~\ref{sec:fast}--\ref{sec:closed2}. Here we keep only the fast coherent
oscillation of the Hubble function. Section~\ref{sec:fast} first recalls field oscillations and particle production in
reheating, and then derives the resonant pair production it drives and the
back-reaction of the pairs, which locks the oscillation in phase with them. Section~\ref{sec:coherent} writes the complete
microscopic system, solves it by a two-timescale expansion, explains the difference between the adiabatic dressing
of the spacetime vacuum and real pairs (with numerical illustrations), and averages over the fast time.
Section~\ref{sec:closed2} gives the closed set of four equations of this case and its relation to
Eq.~\eqref{eq:rate_intro}.

Section~\ref{sec:summary} summarises the results. The longer derivations of the kinetic equation and of the
two-timescale expansion are collected in appendices~\ref{app:kinetic} and~\ref{app:twotime}, and the symbolic checks in
appendix~\ref{app:checks}. Several steps rely on standard tools: adiabatic regularisation~\cite{ParkerFulling1974,AndersonParker1987},
the quantum kinetic (Vlasov) equation for pair production~\cite{KlugerMottolaEisenberg1998,SchmidtEtAl1998},
non-equilibrium quantum field theory~\cite{CalzettaHu}, and the asymptotics of the adiabatic
expansion~\cite{Berry1989,DabrowskiDunne2014,DabrowskiDunne2016}.

\paragraph{Conventions.} Signature $(-,+,+,+)$; natural units $c=\hbar=k_B=1$; $G=\Mpl^{-2}$,
$\mpl=(8\pi G)^{-1/2}=2.43\times10^{18}$\,GeV. Overdots denote cosmic-time derivatives, $\nabla$ is the gradient with
respect to comoving coordinates, and $R=6(\dot H+2H^2)$. Mass dimensions: $[\Phi]=1$, $[H]=[\GaM]=1$, $[\rho]=4$. The symbol $\epsilon$ always denotes the $\epsilon$-rate $-\dot H/H^2$ of
Eq.~\eqref{eq:eps}. The adiabaticity parameter of a mode, $\zeta_k=|\dot\Omega_k|/\Omega_k^2$ (section~\ref{sec:rhoH}), is a different quantity.

\part*{Part I: the adiabatic case}
\addcontentsline{toc}{part}{Part I: the adiabatic case}

\section{Friedmann equations with a time-varying cosmological term}\label{sec:frw}

We start from the Einstein equations with a cosmological term that is allowed to depend on time~\cite{Xue2015}. The equations read
\begin{equation}
  G_{\mu\nu}+\Lambda(t)\,g_{\mu\nu}=8\pi G\,T^{(M+R)}_{\mu\nu},
  \label{eq:einstein}
\end{equation}
where $T^{(M+R)}_{\mu\nu}$ describes the massive pairs and the radiation. When the $\Lambda$ term is moved to the right-hand side,
it behaves as a perfect fluid with energy density $\rho_\Lambda=\Lambda/8\pi G$ and pressure $p_\Lambda=-\rho_\Lambda$. The pairs are a
perfect fluid with $p_M=w_M\rho_M$, where $w_M\ll1$ for massive pairs; the
radiation has $p_R=\rho_R/3$.

For the flat Friedmann metric $\dd s^2=-\dd t^2+a^2(t)\delta_{ij}\dd x^i\dd x^j$, the only non-vanishing Christoffel symbols
are $\Gamma^0_{ij}=a\dot a\,\delta_{ij}$ and $\Gamma^i_{0j}=\Gamma^i_{j0}=H\delta^i_j$, with $H=\dot a/a$. From them one obtains the Ricci tensor,
its trace and the Einstein tensor,
\begin{equation}
  \begin{gathered}
  R_{00}=-3(\dot H+H^2),\qquad R_{ij}=a^2(\dot H+3H^2)\delta_{ij},\qquad R=6(\dot H+2H^2),\\
  G_{00}=3H^2,\qquad G_{ij}=-a^2(2\dot H+3H^2)\delta_{ij}
  \end{gathered}
  \label{eq:geometry}
\end{equation}
(verified with computer algebra, appendix~\ref{app:checks}). The $00$ component of \eqref{eq:einstein}, with
$T^{\rm tot}_{00}=\rho_\Lambda+\rho_M+\rho_R$ and $8\pi G=\mpl^{-2}$, is the first Friedmann equation. The $ij$ component reads
$-(2\dot H+3H^2)=\mpl^{-2}(p_\Lambda+p_M+p_R)$. Adding three times the first equation eliminates $H^2$ and gives
$-2\dot H=\mpl^{-2}\sum_i(\rho_i+p_i)$, in which the cosmological term cancels because $\rho_\Lambda+p_\Lambda=0$. Hence
\begin{align}
  H^2&=\frac{1}{3\mpl^2}\big(\rho_\Lambda+\rho_M+\rho_R\big),\label{eq:F1}\\
  \dot H&=-\frac{1}{2\mpl^2}\Big[(1+w_M)\rho_M+\tfrac43\rho_R\Big].\label{eq:F2}
\end{align}
Only matter decelerates the expansion. It is convenient to introduce the $\epsilon$-rate,
\begin{equation}
  \epsilon\equiv-\frac{\dot H}{H^2}=\frac32\,\frac{(1+w_M)\rho_M+\frac43\rho_R}{\rho_\Lambda+\rho_M+\rho_R},
  \label{eq:eps}
\end{equation}
which approaches $0$ when $\rho_\Lambda$ dominates, equals $\tfrac32$ when the pairs dominate and $2$ when radiation dominates.

The contracted Bianchi identity $\nabla^\mu G_{\mu\nu}\equiv0$ requires the \emph{total} energy--momentum tensor, including the
cosmological term, to be conserved. For perfect fluids the $\nu=0$ component of $\nabla^\mu T^{\rm tot}_{\mu\nu}=0$ is
$\sum_i[\dot\rho_i+3H(\rho_i+p_i)]=0$, that is,
\begin{equation}
  \dot\rho_\Lambda+\big[\dot\rho_M+3H(1+w_M)\rho_M\big]+\big[\dot\rho_R+4H\rho_R\big]=0 .
  \label{eq:bianchi}
\end{equation}
The same result follows by differentiating \eqref{eq:F1} and using \eqref{eq:F2}. For a constant $\Lambda$ each component is
conserved separately. A time-varying cosmological term, by contrast, is consistent only if it exchanges energy with
matter, at the rate
\begin{equation}
  \mathcal Q\equiv-\dot\rho_\Lambda=\big[\dot\rho_M+3H(1+w_M)\rho_M\big]+\big[\dot\rho_R+4H\rho_R\big].
  \label{eq:Q}
\end{equation}
In the scenario of Refs.~\cite{Xue2015,Xue2023,Xue2023b,Xue2024} this energy goes into gravitational pair production. In this article
we derive $\mathcal Q$ from a field theory in two cases. In Part~I it takes the form of the relaxation term of the cosmic rate equation
[Eq.~\eqref{eq:QI}]. In Part~II we find that the fast oscillation of the Hubble function, when generated by the pairs themselves,
exchanges no net energy with them (section~\ref{sec:backreact}). The energy of the created pairs is supplied by the slow decrease of
$\rho_\Lambda$ in the time evolution, and section~\ref{sec:averaged} derives it explicitly as
$\mathcal Q=-\dot\rho_\Lambda=[(3-2\epsilon)\bar H+\Gde]\rhoMH$ [Eq.~\eqref{eq:rate_coh}].

\section{Complex scalar field in the Friedmann universe}\label{sec:field}

We describe the massive pairs by a complex scalar field with action
\begin{equation}
  S_\Phi=\int\dd^4x\sqrt{-g}\,\Big[-g^{\mu\nu}\partial_\mu\Phi^*\partial_\nu\Phi-m^2|\Phi|^2-\xi R\,|\Phi|^2\Big],
  \label{eq:SPhi}
\end{equation}
where $\xi$ is the non-minimal coupling to curvature. The action is invariant under the global U(1) transformation
$\Phi\to e^{i\alpha}\Phi$. Quanta of $\Phi$ are particles $F$ and quanta of $\Phi^*$ are antiparticles $\bar F$, and every process
considered in this article creates or destroys them in pairs. The field equation is
$\ddot\Phi+3H\dot\Phi-a^{-2}\nabla^2\Phi+(m^2+\xi R)\Phi=0$.

\paragraph{Rescaled field.} The Hubble friction $3H\dot\Phi$ can be removed by the rescaling $\chi\equiv a^{3/2}\Phi$. Since
$a^3|\dot\Phi|^2=|\dot\chi|^2-\tfrac32H\,\partial_t|\chi|^2+\tfrac94H^2|\chi|^2$, and the middle term equals
$\tfrac32\dot H|\chi|^2$ up to a total time derivative, the action becomes
\begin{equation}
  S_\Phi=\int\dd t\,\dd^3x\,\Big[|\dot\chi|^2-a^{-2}|\nabla\chi|^2-m^2|\chi|^2-\Delta(t)\,|\chi|^2\Big],\qquad
  \Delta\equiv\xi R-\tfrac94H^2-\tfrac32\dot H .
  \label{eq:Schi}
\end{equation}
The rescaled field is a collection of oscillators of mass $m$, and \emph{its entire coupling to the Friedmann background is
the pairwise term $-\Delta(t)|\chi|^2$}. Because $|\chi|^2$ contains the products $a^\dagger b^\dagger$ and $ab$ of particle and
antiparticle operators, this coupling can only create or annihilate pairs. Each comoving mode obeys
$\ddot\chi_k+\Omega_k^2\chi_k=0$, with
\begin{equation}
  \Omega_k^2(t)=\frac{k^2}{a^2}+m^2+\Delta(t).
  \label{eq:Omega}
\end{equation}

\paragraph{Bogoliubov coefficients.} We expand
$\chi=\int\frac{\dd^3k}{(2\pi)^3}[a_\kk\chi_k(t)+b^\dagger_{-\kk}\chi_k^*(t)]e^{i\kk\cdot\xx}$, with the usual commutators and the Wronskian
normalisation $\chi_k\dot\chi_k^*-\chi_k^*\dot\chi_k=i$. At each instant the mode function can be written on the basis of the
instantaneous positive- and negative-frequency solutions,
$\chi_k=(2\Omega_k)^{-1/2}[\alpha_ke^{-i\theta_k}+\beta_ke^{+i\theta_k}]$ with $\theta_k=\int^t\Omega_k$ and $|\alpha_k|^2-|\beta_k|^2=1$. The mode equation is
then equivalent to the Bogoliubov equations $\dot\alpha_k=(\dot\Omega_k/2\Omega_k)e^{2i\theta_k}\beta_k$ and
$\dot\beta_k=(\dot\Omega_k/2\Omega_k)e^{-2i\theta_k}\alpha_k$~\cite{Parker1969,BirrellDavies}: a changing frequency mixes positive and negative
frequencies, which is pair creation.

\paragraph{Kinetic equations.} Two combinations of the Bogoliubov coefficients carry all the physics: the occupation
$f_k=|\beta_k|^2$, which is the same for particles in $\kk$ and antiparticles in $-\kk$, and the pair coherence
$c_k=\alpha_k\beta_k^*e^{-2i\theta_k}$, which measures the phase correlation between the created particle and antiparticle.
Differentiating these definitions and using the Bogoliubov equations together with $|\alpha_k|^2=1+f_k$, one finds the exact
equations
\begin{equation}
  \dot f_k=\frac{\dot\Omega_k}{\Omega_k}\,\mathrm{Re}\,c_k,\qquad
  \dot c_k=-2i\Omega_k\,c_k+\frac{\dot\Omega_k}{2\Omega_k}\,(1+2f_k).
  \label{eq:kin}
\end{equation}
These are the quantum kinetic (Vlasov) equations of pair production~\cite{KlugerMottolaEisenberg1998,SchmidtEtAl1998}, here for a scalar
in a Friedmann universe. The coherence rotates at twice the mode frequency, and pairs are created only when the change
of frequency $\dot\Omega_k$ acts on it. The equations conserve $|c_k|^2-f_k(1+f_k)$, which vanishes for the pure state evolved
from the spacetime vacuum.

\paragraph{Density and pressure.} The expectation values $\langle|\chi_k|^2\rangle=(1+2f_k+2\mathrm{Re}\,c_k)/2\Omega_k$ and
$\langle|\dot\chi_k|^2\rangle=\Omega_k(1+2f_k-2\mathrm{Re}\,c_k)/2$, inserted into $\rho=|\dot\Phi|^2+a^{-2}|\nabla\Phi|^2+m^2|\Phi|^2$ and
$p=|\dot\Phi|^2-\tfrac13a^{-2}|\nabla\Phi|^2-m^2|\Phi|^2$, give the density and pressure of the pairs. With $g_\Phi$ internal degrees of
freedom, and after the zero-point terms have been absorbed into $\rho_\Lambda$,
\begin{equation}
  \rho_M=\frac{2g_\Phi}{a^3}\int\!\frac{\dd^3k}{(2\pi)^3}\,\Omega_kf_k,\qquad
  p_M=\frac{2g_\Phi}{a^3}\int\!\frac{\dd^3k}{(2\pi)^3}\Big[-\Omega_k\,\mathrm{Re}\,c_k+\frac{p^2}{3\Omega_k}\big(f_k+\mathrm{Re}\,c_k\big)\Big],
  \label{eq:rhop}
\end{equation}
with $p=k/a$ the physical momentum. The density is set by the occupation alone. The pressure also contains a term
carried by the pair coherence, which oscillates at frequency $2\Omega_k$. It averages to zero in Part~I but drives the
dynamics of Part~II.

\section{Pair production and decay: the cosmic rate equation}\label{sec:prod}

\subsection{Adiabatic occupation and the density \texorpdfstring{$\rhoMH$}{rhoMH}}\label{sec:rhoH}

For $m\gg H$ the background changes slowly compared with the mode frequency: the adiabaticity parameter
$\zeta_k=|\dot\Omega_k|/\Omega_k^2$ (not to be confused with the $\epsilon$-rate) is at most of order $H/m$. Equations \eqref{eq:kin} can then be solved order by order in
$\zeta_k$. To first order, dropping $\dot c_k$ in the second equation gives a slaved coherence
$c_k\simeq-i\dot\Omega_k(1+2f_k)/4\Omega_k^2$. It is purely imaginary, so it does not change $f_k$. To second order, the time
derivative of this coherence produces a real part
$\mathrm{Re}\,c_k\simeq(2\Omega_k)^{-1}\dd_t[\dot\Omega_k(1+2f_k)/4\Omega_k^2]$. The first equation then gives
$\dot f_k=(1+2f_k)\,\dd_t(\dot\Omega_k/4\Omega_k^2)^2$, which integrates, for the spacetime vacuum as initial state, to the adiabatic occupation
\begin{equation}
  f_k^{\rm ad}=\Big(\frac{\dot\Omega_k}{4\Omega_k^2}\Big)^2 .
  \label{eq:fad}
\end{equation}
The time dependence of $\Omega_k$ has two sources: the redshift of the physical momentum, and the slow variation of $\Delta$. With
constant $\epsilon$ one finds $\dot\Omega_k=-(H/\Omega_k)[p^2+\epsilon c_\xi H^2]$, with $c_\xi=(12\xi-\tfrac94)-(6\xi-\tfrac32)\epsilon$. For $p\gg H$ the
redshift dominates, and $f_k^{\rm ad}\simeq H^2p^4/16\Omega_k^6$, up to corrections of relative order $H^2/p^2$.

The massive pairs, with equation of state $w_M\simeq0$, are the non-relativistic modes. We therefore integrate the energy
$\Omega_kf_k^{\rm ad}$ of each mode over the shell $p\le\kappa m$, counting particles and antiparticles:
\begin{equation}
  \rhoMH=\frac{g_\Phi H^2}{16\pi^2}\int_0^{\kappa m}\!\frac{p^6\,\dd p}{(p^2+m^2)^{5/2}}
  =2\chi\,m^2H^2,\qquad \chi=\frac{g_\Phi I(\kappa)}{32\pi^2},\qquad
  I(\kappa)=\int_0^\kappa\!\frac{x^6\,\dd x}{(1+x^2)^{5/2}} .
  \label{eq:rhoH}
\end{equation}
Numerically $I(1)=3.57\times10^{-2}$, so $\chi=1.13\times10^{-4}g_\Phi$. The value $\chi\simeq1.85\times10^{-3}$ of
Refs.~\cite{Xue2023,Xue2023b,Xue2024} thus corresponds to $g_\Phi\simeq16$ for $\kappa=1$. The field theory reproduces the scaling
$\rhoMH\propto m^2H^2$ of the scenario and fixes its coefficient in terms of the degeneracy and the non-relativistic cut-off.

Two remarks are needed. First, $\rhoMH$ is a power of $H/m$: it is the adiabatic \emph{dressing} of the spacetime vacuum, which follows the
expansion instantaneously, and it should be distinguished from the creation of real pairs, which a smooth expansion
suppresses as $e^{-2\pi m/H}$. Section~\ref{sec:dressing} explains and illustrates this distinction. Second, if the cut-off is
removed, $I(\kappa)=\tfrac12\kappa^2-\tfrac52\ln\kappa+0.850+\Ocal(\kappa^{-2})$: the relativistic modes add terms $\propto H^2p_c^2$ and
$\propto m^2H^2\ln(p_c/m)$. These are of second adiabatic order, and in a covariant regularisation they are absorbed into
Newton's constant~\cite{ParkerFulling1974,AndersonParker1987}. Retaining the non-relativistic shell as physical pairs is
therefore a choice of the scenario, and it is this choice that fixes $\chi$.

\subsection{Pairwise coupling to the spacetime vacuum: kinetic equation and $\texorpdfstring{\GaM}{GammaM}$}\label{sec:kinetic}

The classical background alone gives the instantaneous occupation \eqref{eq:fad}; it does not make the pair density
relax towards a value. Relaxation requires an irreversible exchange with degrees of freedom of spacetime beyond the
classical background, the ``spacetime vacuum'' of Refs.~\cite{Xue2023,Xue2023b,Xue2024}. At the lowest dimension, they couple to the pairs
through a term of the same U(1)-invariant form as in \eqref{eq:Schi}:
\begin{equation}
  \mathcal L_{\rm S}=-\big[\Delta(t)+\delta\Ocal_S(t,\xx)\big]|\chi|^2 ,
  \label{eq:La}
\end{equation}
where $\delta\Ocal_S$ is a fluctuating operator of mass dimension 2 with zero mean. Only its spectral function
$S_{\mathcal O}(\omega)=\int\dd\tau\,e^{i\omega\tau}\langle\delta\Ocal_S(\tau)\delta\Ocal_S(0)\rangle$ enters at the order we need.

To second order in $\delta\Ocal_S$, Fermi's golden rule gives the rate at which a pair $F(\kk)\bar F(-\kk)$ is created,
$\Gamma^\uparrow_k=S_{\mathcal O}(-2\Omega_k)/4\Omega_k^2$, and the coefficient of pair annihilation,
$\Gamma^\downarrow_k=S_{\mathcal O}(+2\Omega_k)/4\Omega_k^2$. The sign of the frequency reflects the fact that the spacetime vacuum must supply the
energy $2\Omega_k$ to create a pair, and absorb it to annihilate one. With Bose enhancement and for a random-phase ensemble,
the occupations obey (appendix~\ref{app:kinetic})
\begin{equation}
  \dot f_F(\kk)=\dot f_{\bar F}(-\kk)=\Gamma^\uparrow_k\,[1+f_F][1+f_{\bar F}]-\Gamma^\downarrow_k\,f_Ff_{\bar F} .
  \label{eq:Boltz}
\end{equation}
Particles and antiparticles change together, as they must for a pairwise process. The stationary occupation is
$f^*_k=[(\Gamma^\downarrow_k/\Gamma^\uparrow_k)^{1/2}-1]^{-1}$. Linearising \eqref{eq:Boltz} around it and integrating over the modes with weight
$\Omega_k$ gives the relaxation term of the rate equation:
\begin{equation}
  \dot\rho_M+3H\rho_M=\GaM\big(\rhoMH-\rho_M\big),\qquad
  \GaM=\frac{2\Gamma^\uparrow(1+f^*)}{f^*}\simeq2\big(\Gamma^\uparrow\Gamma^\downarrow\big)^{1/2}\ \ (f^*\ll1),
  \label{eq:GaM}
\end{equation}
where $\rhoMH$ is now the energy density of the stationary occupation $f^*$. $\GaM$ is thus the \emph{relaxation rate} of the pair
density. For dilute pairs the full kinetics gives
$\dot\rho_M+3H\rho_M=\tfrac12\GaM\rhoMH-\tfrac12\GaM\rho_M^2/\rhoMH$: production proceeds at $\tfrac12\GaM$ per equilibrium pair and
annihilation at $\tfrac12\GaM\rho_M/\rhoMH$ per pair. Their difference reduces to \eqref{eq:GaM} at linear order in $\rhoMH-\rho_M$,
which corresponds to equal production and annihilation rates. In the scenario of
Refs.~\cite{Xue2023,Xue2023b,Xue2025} the rate is estimated as $\GaM\simeq\chi m\epsilon/4\pi$; its microscopic value depends on the spectrum
of the fluctuations of the spacetime vacuum (appendix~\ref{app:kinetic}).

\subsection{Decay into light fermions}\label{sec:decay}

During reheating the massive pairs must decay into relativistic particles. We describe this with light fermions: a
Dirac fermion $\psi$ carrying U(1) charge $+1$ and a neutral fermion $N$. The renormalisable coupling of $\Phi$ to them that
respects the global U(1) symmetry is the Yukawa interaction
\begin{equation}
  \mathcal L_Y=-\,y\left(\Phi\,\bar\psi N+\Phi^*\,\bar N\psi\right),\qquad [y]=0 .
  \label{eq:LY}
\end{equation}
The field $\Phi$ lowers the charge by one unit (it annihilates $F$ or creates $\bar F$) and $\bar\psi$ raises it by one, so each term
is neutral. The first term mediates $F\to\psi\bar N$ and the second $\bar F\to\bar\psi N$. A coupling of the form
$(\Phi+\Phi^*)\bar\psi\psi$, by contrast, would break the U(1) symmetry of the action, and we exclude it.

The decay width follows from the tree-level amplitude $\mathcal M=y\,\bar u(p_\psi)v(p_N)$. Summing over spins,
$\sum|\mathcal M|^2=y^2\,\mathrm{Tr}[(\gamma\cdot p_\psi)(\gamma\cdot p_N)]=4y^2\,p_\psi\cdot p_N$. For massless final states $2p_\psi\cdot p_N=m^2$, so
$\sum|\mathcal M|^2=2y^2m^2$. The two-body phase space of massless particles gives $\Gamma=\sum|\mathcal M|^2/(16\pi m)$, hence
\begin{equation}
  \Gde=\frac{y^2m}{8\pi}\equiv g_Y^2\,m ,\qquad g_Y^2\equiv\frac{y^2}{8\pi}.
  \label{eq:Gde}
\end{equation}
By CPT invariance $\Gamma(F\to\psi\bar N)=\Gamma(\bar F\to\bar\psi N)$, so
particles and antiparticles decay at the same rate. A pair in a mode of energy $\Omega_k$ decays at the time-dilated rate
$\Gde\,m/\Omega_k\simeq\Gde$ for non-relativistic pairs, so the decays add $-\Gde f_{F,\bar F}$ to the kinetic equation \eqref{eq:Boltz}.
Inverse decays, $\psi\bar N\to F$, are negligible as long as the decay products are much lighter and colder than $m$. The energy
released, $\Gde\rho_M$ per unit volume and time, goes into the relativistic $\psi$ and $N$, i.e.\ into radiation.

\subsection{Complete effective Lagrangian and the cosmic rate equation}\label{sec:Leff}

Collecting the pieces, the effective Lagrangian of the model is
\begin{align}
  \mathcal L_{\rm eff}=\sqrt{-g}\,\Big[&-|\partial\Phi|^2-m^2|\Phi|^2-\xi R|\Phi|^2-\delta\Ocal_S|\Phi|^2\nonumber\\
  &-\bar\psi\gamma^\mu\nabla_\mu\psi-\bar N\gamma^\mu\nabla_\mu N-y\big(\Phi\bar\psi N+\Phi^*\bar N\psi\big)\Big]+\mathcal L_{\mathcal S}.
  \label{eq:Leff}
\end{align}
The first three terms describe the pairs and their coupling to the classical background. In terms of $\chi$ they are the
action \eqref{eq:Schi}, and together with the fourth term they give the pairwise coupling \eqref{eq:La} to the spacetime vacuum. The
next three terms describe the decay products and the Yukawa decay. $\mathcal L_{\mathcal S}$ stands for the dynamics of the spacetime
vacuum, which enters only through the spectral function of $\delta\Ocal_S$. All terms are invariant under the global U(1)
symmetry, and all have mass dimension four.

Adding the decay term to the kinetic equation \eqref{eq:Boltz} and integrating with weight $\Omega_k$, as in
section~\ref{sec:kinetic}, gives the cosmic rate equation, and the energy of the decays gives the radiation equation:
\begin{equation}
  \boxed{\;\dot\rho_M+3H\rho_M=\GaM\big(\rhoMH-\rho_M\big)-\Gde\,\rho_M,\qquad
  \dot\rho_R+4H\rho_R=\Gde\,\rho_M\;}
  \label{eq:rate}
\end{equation}
These are Eq.~\eqref{eq:rate_intro} and the reheating equation of Refs.~\cite{Xue2023b,Xue2025}. Inserting them into the energy
balance \eqref{eq:bianchi} determines the energy transfer from the cosmological term,
\begin{equation}
  \mathcal Q=-\dot\rho_\Lambda=\GaM\big(\rhoMH-\rho_M\big):
  \label{eq:QI}
\end{equation}
pairs are produced at the expense of the cosmological term when $\rho_M<\rhoMH$, and return energy to it when $\rho_M>\rhoMH$.

\section{The closed set of background equations (adiabatic case)}\label{sec:closed1}

\subsection{The closed set}

The Friedmann equations \eqref{eq:F1}--\eqref{eq:F2}, the rate equation and the radiation equation \eqref{eq:rate}, and the
energy transfer \eqref{eq:QI} form a closed system for the four unknowns $(H,\rho_\Lambda,\rho_M,\rho_R)$. The two Friedmann equations, the rate
equation and the radiation equation are four independent equations that determine the four unknowns. The energy transfer
\eqref{eq:QI} then follows from them through the Bianchi identity. To write the system in dimensionless form, we take the pair mass
as the mass scale, $m_*=m$, and choose a reference epoch with Hubble function $H_*$ and scale factor $a_*$. These characterise a specific stage of the Universe's evolution: for
example, the end of inflation when reheating is studied, the moment at which the pivot scale of the cosmic microwave
background leaves the horizon when inflation is studied, or the present epoch, $H_*=H_0$, when late-time cosmology is
studied. With the e-folding variable $x=\ln(a/a_*)$ and the normalised quantities $h=H/H_*$ and
$\Omega_i=\rho_i/(3\mpl^2H_*^2)$, the closed set of four equations is~\cite{Xue2023b}
\begin{equation}
  \boxed{
  \begin{gathered}
  h^2=\Omega_\Lambda+\Omega_M+\Omega_R,\qquad \frac{\dd h^2}{\dd x}=-3\Omega_M-4\Omega_R,\\[2pt]
  \frac{\dd\Omega_M}{\dd x}+3\Omega_M=\frac{\GaM}{H}\big(\Omega_M^H-\Omega_M\big)-\frac{\Gde}{H}\Omega_M,\qquad
  \frac{\dd\Omega_R}{\dd x}+4\Omega_R=\frac{\Gde}{H}\Omega_M,
  \end{gathered}}
  \label{eq:closed1}
\end{equation}
with $\Omega_M^H=\tfrac23\chi(m/\mpl)^2h^2$ and $w_M$ neglected. The derivation has fixed the meaning of each input. The pair mass $m$
and the coefficient $\chi=g_\Phi I(\kappa)/32\pi^2$ come from the adiabatic occupation. The relaxation rate $\GaM$ comes from the
spectral function of the fluctuations of the spacetime vacuum [Eq.~\eqref{eq:GaM}]. The decay rate $\Gde=y^2m/8\pi$ comes from
the Yukawa coupling [Eq.~\eqref{eq:Gde}]. Given these inputs and the initial values at the reference epoch,
Eqs.~\eqref{eq:closed1} determine the evolution of the background.

\subsection{Exact solution in the detailed-balance regime}\label{sec:exact}

When the relaxation is fast, the rate equation \eqref{eq:rate} can be solved by iteration in $H/\GaM$ and $\Gde/\GaM$:
\begin{equation}
  \rho_M=\rhoMH-\frac{(3-2\epsilon)H+\Gde}{\GaM}\,\rhoMH+\Ocal\Big(\frac{H^2}{\GaM^2}\Big)
  \qquad\Longrightarrow\qquad
  \GaM\gg H,\ \Gde:\quad \rho_M\simeq\rhoMH .
  \label{eq:detbal}
\end{equation}
The pair density is then pinned to $\rhoMH$, and the production and annihilation of pairs balance each other. This is the
detailed balance of Ref.~\cite{Xue2023b}, where the back-and-forth exchange averages to $\langle\rho_M-\rhoMH\rangle=0$ over its period, and
which characterises the pair-dominated episode, $\GaM>H>\Gde$. If, in addition, decays and radiation can be neglected, the system can be solved exactly. Inserting
$\rho_M=2\chi m^2H^2$ into the $\dot H$ equation gives $\dot H=-\chi(m/\mpl)^2H^2$, i.e.\ a \emph{constant} $\epsilon$-rate, and then
\begin{equation}
  \epsilon=\chi\,\frac{m^2}{\mpl^2},\qquad H=\frac{1}{\epsilon t},\qquad a\propto t^{1/\epsilon},\qquad
  \rho_\Lambda=3\mpl^2H^2\Big(1-\frac{2\epsilon}{3}\Big),\qquad
  \mathcal Q=(3-2\epsilon)\,H\,\rhoMH
  \label{eq:exact}
\end{equation}
(verified symbolically). In this regime the source $\GaM(\rhoMH-\rho_M)$ of the rate equation tends to the finite value
$(\partial_t+3H)\rhoMH=(3-2\epsilon)H\rhoMH$: the power needed to keep the pairs on $\rhoMH\propto H^2$ while the Universe expands. The
cosmological term decays as $H^2$ and feeds the pairs.
The pair fraction $\rho_M/3\mpl^2H^2=\tfrac23\epsilon$ is fixed by the mass alone. Two regimes follow.
\begin{itemize}[leftmargin=1.5em,itemsep=2pt]
  \item For $\chi m^2<\mpl^2$, i.e.\ $\epsilon<1$, the expansion accelerates: the solution describes power-law inflation driven by the
  cosmological term and slowed down by the pairs.
  \item Pair domination ($\epsilon=\tfrac32$, $\rho_\Lambda=0$) requires $m=(3/2\chi)^{1/2}\mpl=28.5\,\mpl$ for $\chi=1.85\times10^{-3}$. This is close
  to the value $27.7\,\mpl$ used in Ref.~\cite{Xue2025}, for which \eqref{eq:exact} gives $\epsilon=1.42$.
\end{itemize}
The solution requires $\rhoMH\le3\mpl^2H^2-\rho_R$, i.e.\ $\rho_\Lambda\ge0$. The same solution is recovered in Part~II from the fast
coherent dynamics (section~\ref{sec:closed2}).

\subsection{Applications: inflation and standard cosmology}\label{sec:applications}

\paragraph{Inflation.} During inflation the cosmological term dominates, $\Omega_\Lambda\simeq h^2$, and $\epsilon\ll1$. The natural reference
epoch is the moment at which the scales observed in the cosmic microwave background leave the horizon. Pair production
then acts as a brake. The pairs, held near $\rhoMH\propto H^2$, give the expansion a non-zero $\epsilon$-rate and convert part of the
energy of the cosmological term into matter. In the detailed-balance solution \eqref{eq:exact} the brake is constant, $\epsilon=\chi m^2/\mpl^2$.
The end of inflation therefore requires the departure from detailed balance described by the finite-$\Gamma_M$ terms of
\eqref{eq:closed1}, when $\Gamma_M$ becomes comparable to $H$. In Ref.~\cite{Xue2023} inflation driven by the time-varying cosmological term and slowed down by the massive pair
plasma was studied in this framework. The resulting relation between the spectral index and the tensor-to-scalar ratio is
consistent with observations, and a singularity-free pre-inflationary phase was discussed. The subsequent episodes of
preheating, pair domination and reheating follow from the numerical solution of \eqref{eq:closed1}~\cite{Xue2023b}.

\paragraph{Standard cosmology after reheating.} After reheating the Hubble function decreases by many orders of magnitude, and
with it $\rhoMH=2\chi m^2H^2$. Pairs that are stable on cosmological time scales ($\Gde\simeq0$) behave as cold dark
matter~\cite{Xue2023b}, and their density soon exceeds $\rhoMH$. The energy transfer \eqref{eq:QI} then changes sign,
$\mathcal Q=\GaM(\rhoMH-\rho_M)<0$, and energy flows back from matter to the cosmological term. In Ref.~\cite{Xue2024} a closed set of
equations of this type was solved with initial conditions given by observations, taking the present epoch as reference
($H_*=H_0$). Dark energy decreases into matter and radiation during inflation and reheating. In standard cosmology, matter
and radiation convert back into dark energy until the present epoch. These results were proposed as possible solutions of
the cosmological fine-tuning and coincidence problems~\cite{Xue2024}. The closed set \eqref{eq:closed1} derived here contains the
matter channel of this exchange. A direct exchange between the cosmological term and radiation would require an additional
coupling, which we have not derived.

\subsection{Why is $\texorpdfstring{\GaM}{GammaM}$ proportional to the pair mass?}\label{sec:whym}

The kinetic derivation of section~\ref{sec:kinetic} identifies $\GaM$ as the relaxation rate of the pair density, but it does not fix
its value, which depends on the spectral function of the spacetime-vacuum fluctuations. In Refs.~\cite{Xue2015,Xue2023,Xue2023b,Xue2024} the rate is estimated as $\GaM\simeq\chi m\epsilon/4\pi$, proportional to the pair mass and independent of $H$. Three arguments make this scaling
plausible.

\emph{(i) Counting pairs within the horizon.} The number of pairs in a Hubble volume is
$N\simeq\nMH H^{-3}/2=\chi m/2H$, with $\nMH=\rhoMH/2m=\chi mH^2$. Up to the factor $\chi$, $N$ is the number of Compton
wavelengths $m^{-1}$ that fit in the Hubble radius $H^{-1}$. As the horizon grows, $\dd(H^{-1})/\dd t=\epsilon$, new pairs enter at the rate
$\dot N=\chi m\epsilon/2$. This rate depends on $m$ but not on $H$, and $\GaM=\dot N/2\pi=\chi m\epsilon/4\pi$ is the estimate quoted above. It
vanishes in exact de Sitter space ($\epsilon=0$), where the horizon does not grow.

\emph{(ii) The microscopic time scale.} Creation and annihilation of a pair of mass $m$ are quantum processes on the Compton
time $m^{-1}$. In the fast coherent dynamics of Part~II they appear as a back-and-forth oscillation at frequency $2m$
(section~\ref{sec:split}). Written as $\GaM=(\chi\epsilon/8\pi)\times2m$, the rate is a small fraction, set by the pair abundance $\chi$ and the
expansion rate $\epsilon$, of this microscopic frequency.

\emph{(iii) Scale-free fluctuations of the spacetime vacuum.} By section~\ref{sec:kinetic}, the production and annihilation
coefficients are $\Gamma^{\uparrow,\downarrow}=S_{\mathcal O}(\mp2m)/4m^2$, where $S_{\mathcal O}$ has mass dimension 3. If the fluctuations of the spacetime vacuum have no intrinsic
scale at the frequency $2m$, dimensional analysis requires $S_{\mathcal O}(\omega)=\sigma_\pm|\omega|^3$, with dimensionless $\sigma_\pm$ for positive and
negative frequencies. Then $\Gamma^{\uparrow,\downarrow}=2\sigma_\mp m$, and
\begin{equation}
  \GaM=2\big(\Gamma^\uparrow\Gamma^\downarrow\big)^{1/2}=4\,(\sigma_+\sigma_-)^{1/2}\,m\;\propto\;m .
  \label{eq:Gm}
\end{equation}
The proportionality to $m$ thus follows from scale invariance of the fluctuations at the pair-creation frequency. Comparison
with $\GaM=\chi m\epsilon/4\pi$ would correspond to $(\sigma_+\sigma_-)^{1/2}=\chi\epsilon/16\pi$. These arguments make the scaling plausible but do not
derive the coefficient. Its value requires a microscopic theory of the spacetime vacuum.

\part*{Part II: the fast-oscillating case}
\addcontentsline{toc}{part}{Part II: the fast-oscillating case}

\section{Fast oscillation of the Hubble function and resonant pair production}\label{sec:fast}

\subsection{Field oscillations and particle production in reheating}\label{sec:reheating_intro}

In the standard picture of reheating, the inflaton $\varphi$ oscillates coherently about the minimum of its potential at the
end of inflation, and its energy is transferred to other fields while it oscillates~\cite{AbbottFarhiWise1982,AlbrechtSteinhardtTurnerWilczek1982}.
If the coupling is weak, the transfer proceeds through perturbative decays of the inflaton quanta. If the oscillation amplitude
is large, a much faster channel opens. A field $\tilde\chi$ coupled to the inflaton acquires a time-dependent effective mass,
$m_{\rm eff}^2(t)=m_{\tilde\chi}^2+g^2\varphi^2(t)$, and its modes obey an oscillator equation with a periodically varying frequency of the
Mathieu type. Such an equation has bands of momenta in which the mode amplitudes grow exponentially, and particles in these
bands are produced explosively. This is parametric resonance, or preheating~\cite{DolgovKirilova1990,TraschenBrandenberger1990,KofmanLindeStarobinsky1994,ShtanovTraschenBrandenberger1995,KofmanLindeStarobinsky1997}.
For small amplitudes the resonance is narrow: only modes whose frequency is close to half the frequency of the oscillating
term, or a multiple of it, grow, with a rate proportional to the amplitude. For large amplitudes, the resonance is broad, and
the growth is stochastic. Reviews of these processes are given in Refs.~\cite{AllahverdiEtAl2010,AminEtAl2015,KolbTurner}.

Even fields that couple to the inflaton only through gravity are affected by its oscillations. The oscillating energy density
and pressure of the inflaton make the Hubble function and the Ricci scalar oscillate at twice the inflaton mass. A field
coupled minimally, or through $\xi R|\tilde\chi|^2$, to this oscillating background is then produced by the same resonance mechanism,
without the exponential suppression of production in a smooth expansion~\cite{EmaJinnoMukaidaNakayama2015,EmaJinnoMukaidaNakayama2015b,EmaJinnoMukaidaNakayama2016,EmaNakayamaTang2018}.
Gravitational particle production of this kind is a source of dark matter and other relics~\cite{KolbLong2024,ChungKolbRiotto1999,ChungCrottyKolbRiotto2001}.

In the scenario of Refs.~\cite{Xue2023,Xue2023b} there is no inflaton field. The oscillation of the Hubble function is generated by the
massive pairs themselves, whose pressure oscillates at twice their mass. The pairs are therefore both the source of the
oscillating background and the field that it produces. In this section we study this self-consistent case: the resonance
that the oscillation drives, and the back-reaction of the pairs that limits it.

\subsection{Fast--slow decomposition}

Earlier results motivate the form of the fast oscillation that we adopt.
\begin{itemize}[leftmargin=1.5em,itemsep=2pt]
  \item Parker and Fulling~\cite{ParkerFulling1973} showed that a quantised massive scalar field in a coherent pair state has an energy--momentum
  tensor with terms oscillating at twice the mass. They come from the interference $\mathrm{Re}(\alpha^*\beta e^{2imt})$ between positive- and
  negative-frequency components, and they give an oscillating, possibly negative, pressure that can act back on the expansion.
  \item In Refs.~\cite{Xue2023,Xue2023b} the quantities that vary on the Hubble time $\Ocal(H^{-1})$ were separated from those that vary on the Compton time
  $\Ocal(m^{-1})$, and the equations for the fast components were solved numerically. The fast parts of the Hubble function, of
  the cosmological term and of the pair density and pressure were found to oscillate coherently, with a single frequency set by
  the pair mass and a small relative amplitude, while their averages over the microscopic time form a massive pair plasma state.
  The back-and-forth production and annihilation of pairs, averaged over its period, was shown to
  amount to detailed balance between the pairs and the spacetime vacuum, $\langle\rho_M-\rhoMH\rangle=0$, which underlies the rate equation.
\end{itemize}
Our own analysis supports the same form. In \eqref{eq:rhop} the only fast time dependence of the non-relativistic pairs is the
pair coherence $c_k$, which rotates at $2\Omega_k\simeq2m$. The pressure that sources the oscillation of $H$ therefore oscillates at $2m$,
with an envelope that varies on the slow time scales $H^{-1}$ and $\GaM^{-1}$. Higher harmonics are suppressed by further powers of the
small amplitude. We therefore write $H=\bar H+\hf{}$, with
\begin{equation}
  \hf{}(t)=\mathrm{Re}\big[\tilde h\,e^{-2imt}\big],\qquad |\tilde h|\equiv\eta\,\bar H ,
  \label{eq:hf}
\end{equation}
where $\tilde h$ varies slowly and $\eta$ is the relative amplitude of the oscillation. Similarly
$\rho_\Lambda=\bar\rho_\Lambda+\rho_\Lambda^{\rm fast}$. To first order in fast quantities, the Friedmann equations \eqref{eq:F1} and \eqref{eq:F2} give
$2\bar H\hf{}=(\rho_\Lambda^{\rm fast}+\rho_M^{\rm fast})/3\mpl^2$ and $\dot h_f^{\rm fast}=-(\rho^{\rm fast}+p^{\rm fast})/2\mpl^2$. We show in
section~\ref{sec:backreact} that the pair density has no component at frequency $2m$. Since moreover $\rho_\Lambda+p_\Lambda=0$, these
reduce to
\begin{equation}
  \dot h_f^{\rm fast}=-\frac{p_M^{\rm fast}}{2\mpl^2},\qquad \rho_\Lambda^{\rm fast}=6\mpl^2\bar H\hf{} .
  \label{eq:fastF}
\end{equation}
The oscillation of $H$ is driven by the oscillating pressure of the pairs, and the cosmological term oscillates in phase
with it.

\subsection{Resonant coupling and production rate}\label{sec:resonance}

\paragraph{The oscillating coupling.} The fast part of $H$ enters the mode equation through $\Delta=\xi R-\tfrac94H^2-\tfrac32\dot H$.
To first order in $\hf{}$, using $R=6(\dot H+2H^2)$,
\begin{equation*}
  \Delta=\bar\Delta+\big(6\xi-\tfrac32\big)\dot h_f^{\rm fast}+\big(24\xi-\tfrac92\big)\bar H\hf{} .
\end{equation*}
Since $\dot h_f^{\rm fast}\sim2m\,\hf{}\gg\bar H\hf{}$, the last term is negligible, and the fast coupling is
$\Delta_f\simeq(6\xi-\tfrac32)\dot h_f^{\rm fast}=\mathrm{Re}[\tilde\Delta e^{-2imt}]$, with $\tilde\Delta=-2im(6\xi-\tfrac32)\tilde h$. Two features stand out. The coupling
oscillates at exactly $2m$, twice the rest energy of a quantum, so it can create a pair at rest resonantly. And its strength
receives two contributions, one from the Hubble friction (through the rescaling of section~\ref{sec:field}) and one from the
curvature coupling $\xi R$.

\paragraph{Two-timescale equations.} Because the fast coupling is resonant with the rest frequency, it is natural to
separate the fast phases of the particle and antiparticle components of each mode,
$\chi_k=(2m)^{-1/2}[e^{-imt}\psi_+(\kk)+e^{imt}\psi_-^*(-\kk)]$, where $\psi_\pm$ vary slowly. Substituting into
$\ddot\chi_k+(\omega_k^2+\Delta)\chi_k=0$ with $\omega_k^2=k^2/a^2+m^2$, neglecting $\ddot\psi_\pm$ and the non-resonant terms
$\propto e^{\pm3imt}$, and collecting the coefficients of $e^{-imt}$ and $e^{+imt}$, we obtain two coupled first-order equations,
\begin{equation}
  \dot\psi_+(\kk)=-i\delta_k\,\psi_+(\kk)+g\,\tilde h\,\psi_-^*(-\kk),\qquad
  \dot\psi_-^*(-\kk)=+i\delta_k\,\psi_-^*(-\kk)+g\,\tilde h^*\,\psi_+(\kk),
  \label{eq:pump}
\end{equation}
where $\delta_k=k^2/2ma^2$ is the kinetic energy of the mode and
\begin{equation}
  g=\frac12\Big(\frac32-6\xi\Big)
  \label{eq:g}
\end{equation}
is the resonant coupling (verified symbolically). The pump $g\tilde h$ couples each particle mode $\kk$ to the conjugate antiparticle mode
$-\kk$, i.e.\ it creates and annihilates pairs. For minimal coupling, $g=\tfrac34$ reproduces the Bogoliubov coupling
$C_0=\tfrac32H$ of Ref.~\cite{Xue2023}. For conformal coupling, $\xi=\tfrac16$, one has $g=\tfrac14$, and for $\xi=\tfrac14$ the resonant coupling
vanishes at this order.

\paragraph{Solution.} The Heisenberg operators $a_\kk(t)$ and $b^\dagger_{-\kk}(t)$ obey the same linear equations as $\psi_+$ and
$\psi_-^*$. Choosing $\tilde h$ real and writing $\Ups=g|\tilde h|=|g|\eta\bar H$, the evolution matrix of \eqref{eq:pump}
squares to $\mu_k^2$ times the identity, with
\begin{equation}
  \mu_k=\big(\Ups^2-\delta_k^2\big)^{1/2}.
  \label{eq:mu}
\end{equation}
Its exponential is therefore $\cosh(\mu_kt)$ times the identity plus $\sinh(\mu_kt)/\mu_k$ times the matrix. Starting from the
spacetime vacuum, the number of particles produced in mode $\kk$ (equal to the number of antiparticles in $-\kk$) is
\begin{equation}
  n_k(t)=\frac{\Ups^2}{\mu_k^2}\,\sinh^2(\mu_kt)\qquad(\delta_k<\Ups),
  \label{eq:nk}
\end{equation}
and $n_k=\Ups^2\sin^2(\nu_kt)/\nu_k^2$, with $\nu_k^2=\delta_k^2-\Ups^2$, otherwise.

\begin{figure}[t]
\centering
\includegraphics[width=\linewidth]{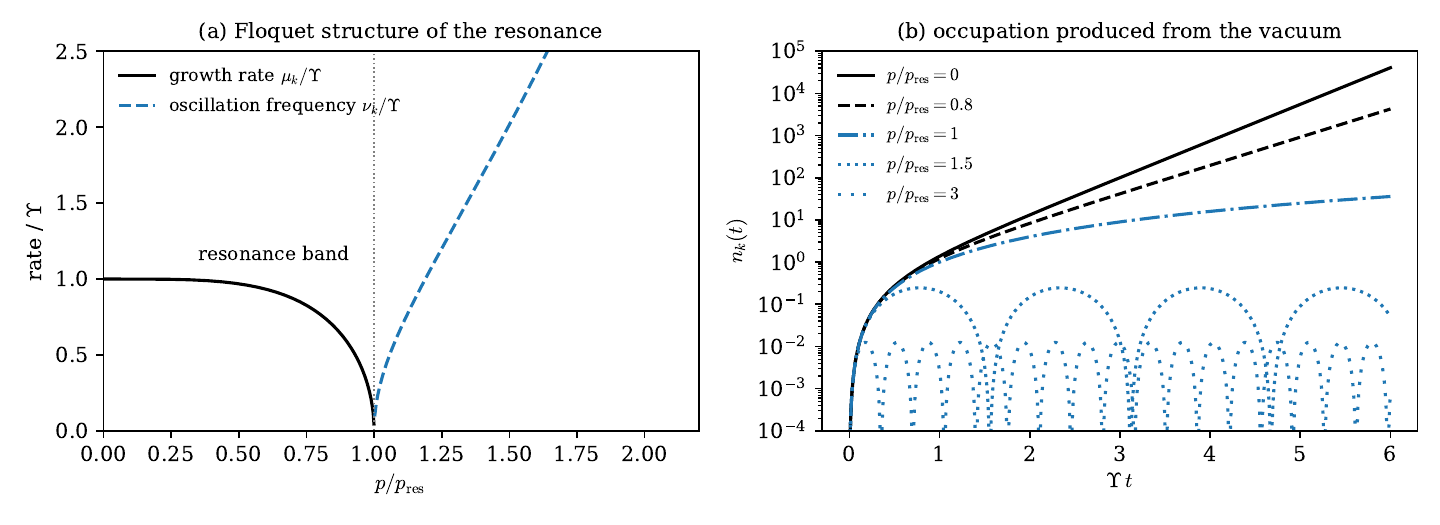}
\caption{Resonant pair production by the oscillating Hubble function. (a)~Growth rate $\mu_k$ inside the resonance band and
oscillation frequency $\nu_k$ outside it, in units of $\Ups=|g|\eta\bar H$, as functions of the physical momentum in units of
$p_{\rm res}=(2m|g|\eta\bar H)^{1/2}$. (b)~Occupation $n_k(t)$ produced from the spacetime vacuum for several momenta.}
\label{fig:resonance}
\end{figure}

\paragraph{The resonance band.} We call \emph{resonance band} the set of modes for which $\mu_k$ is real, i.e.\ whose kinetic energy is
smaller than the pump strength, $\delta_k<\Ups$. In terms of the physical momentum $p=k/a$ it is the range
\begin{equation}
  0\le p<p_{\rm res},\qquad p_{\rm res}\equiv\big(2m\,\Ups\big)^{1/2}=\big(2m|g|\eta\bar H\big)^{1/2}.
  \label{eq:band}
\end{equation}
Inside the band the occupation grows exponentially; outside it ($p>p_{\rm res}$) it only oscillates, with amplitude
$\Ups^2/\nu_k^2$. This is the narrow first band of a Mathieu equation, centred on modes at rest, whose frequency $m$ is half
the pump frequency $2m$. Figure~\ref{fig:resonance} shows the structure. Right at the band edge $\mu_k=0$, and the growth is quadratic,
$n_k=\Ups^2t^2$.

\paragraph{Velocity of the pairs.} The growth rate is largest at $p=0$, where
$\mu_0=\Ups=|g|\eta\bar H$, so that the efficiency per Hubble time is $\mu_0/\bar H=|g|\eta$. The fastest pairs produced have group
velocity $v_{\rm res}=p_{\rm res}/m=(2|g|\eta\bar H/m)^{1/2}$, much smaller than one for $m\gg\bar H$ and $\eta\lesssim1$. The resonance
therefore produces non-relativistic pairs, as required for the $w_M\simeq0$ fluid of Part~I.

\paragraph{Production rate.} Summing \eqref{eq:nk} over the band, the number density is dominated, for $\mu_0t\gg1$, by the
modes near $p=0$. There $\mu_p\simeq\mu_0-p^4/8m^2\mu_0$ and $\sinh^2(\mu_pt)\simeq\tfrac14e^{2\mu_pt}$. Using
$\int_0^\infty p^2e^{-\alpha p^4}\dd p=\Gamma(\tfrac34)/4\alpha^{3/4}$ with $\alpha=t/4m^2\mu_0$, the number density grows as
$n(t)\propto t^{-3/4}e^{2\mu_0t}$, so the production rate per pair is
\begin{equation}
  \Gamma_{\rm prod}\equiv\frac{\dot n}{n}=2\mu_0-\frac{3}{4t}\;\longrightarrow\;2\mu_0=\Big|\frac32-6\xi\Big|\,\eta\,\bar H .
  \label{eq:Gprod}
\end{equation}
At early times, $\mu_0t\ll1$, one has instead $n_0\simeq\mu_0^2t^2$: the Bogoliubov amplitude grows linearly in time.

\paragraph{Why there is no factor $e^{-m/H}$.} For a smooth background, the first-order Bogoliubov amplitude
$\beta_k=\int\dd t\,(\dot\Omega_k/2\Omega_k)e^{-2i\theta_k}$ is the Fourier transform, at frequency $\simeq2m$, of a function that varies on the
time scale $\bar H^{-1}$. If $\dot\Omega_k/\Omega_k$ is analytic in a strip $|\mathrm{Im}\,t|<\tau$ with $\tau\sim\bar H^{-1}$, this transform is bounded
by $e^{-2m\tau}$. For de Sitter space it gives the familiar $|\beta_k|^2\sim e^{-2\pi m/H}$~\cite{BirrellDavies,Mottola1985}. The oscillating
part of the background, by contrast, has a Fourier component \emph{exactly} at $2m$, of amplitude $|\tilde\Delta|=4m|g|\eta\bar H$. The
integrand of $\beta_k$ then has a part that does not oscillate, and $|\beta_k|$ grows first linearly and then exponentially. The
production rate depends on $\bar H$ only through the amplitude $\eta\bar H$ of the oscillation.

\paragraph{Incoherent regime.} The exponential growth \eqref{eq:nk} assumes that the oscillation keeps a fixed phase, so that the
amplitudes created at successive times add coherently. Suppose instead that the phase of the oscillation wanders,
$\tilde h(t)=|\tilde h|e^{i\varphi(t)}$, with $\varphi(t)$ performing a random walk. The \emph{coherence time} $\tau_c$ is defined by the decay of the
phase correlation,
\begin{equation}
  \big\langle e^{i[\varphi(t)-\varphi(t')]}\big\rangle=e^{-|t-t'|/\tau_c},
  \label{eq:tauc}
\end{equation}
i.e.\ $\tau_c$ is the time over which the oscillation remembers its phase. The spectrum of the oscillation is then a Lorentzian of width $\tau_c^{-1}$
centred at $\pm2m$. If $\tau_c$ is short compared with the growth time, $\mu_0\tau_c\ll1$, the pair amplitude created in each interval
$\tau_c$ has a random phase relative to the previous ones. Probabilities, rather than amplitudes, then add. To first order the
Bogoliubov amplitude is $\beta\simeq\int\dd t\,\mu_0e^{i\varphi(t)}$, and with \eqref{eq:tauc}
$\langle|\beta|^2\rangle=\mu_0^2\int\!\!\int\dd t\,\dd t'\,e^{-|t-t'|/\tau_c}\simeq2\mu_0^2\tau_c\,t$ for $t\gg\tau_c$. The occupation thus grows linearly in time,
at the rate
\begin{equation}
  \Gamma^\uparrow_0=\frac{|\tilde\Delta|^2\tau_c}{8m^2}=2\mu_0^2\tau_c ,
  \label{eq:Gincoh}
\end{equation}
which is the golden rule of section~\ref{sec:kinetic} applied to the Lorentzian spectrum. It is smaller than the coherent rate $2\mu_0$ by
the factor $\mu_0\tau_c\ll1$: dephasing turns exponential (parametric) growth into a slow kinetic rate. A
stochastic simulation of \eqref{eq:pump} with a phase-diffusing pump confirms this to within $2\%$. Measuring rates in units of
the coherent growth rate $\mu_0$ (the simulation uses $\mu_0=0.02$ and $\tau_c=2$ and $5$ in units of an arbitrary time scale), we find
$\Gamma^\uparrow_0/\mu_0=0.082$ and $0.196$ for $\mu_0\tau_c=0.04$ and $0.1$, against the predicted $2\mu_0\tau_c=0.080$ and $0.200$.
Because a classical pump has a symmetric spectrum, it gives $\Gamma^\uparrow=\Gamma^\downarrow$. The occupation then grows without reaching a
stationary value, $\dot f=\Gamma^\uparrow(1+2f)$, until back-reaction intervenes.

\subsection{Back-reaction and phase locking}\label{sec:backreact}

\paragraph{Density and pressure of the pair field.} For a homogeneous field with the decomposition above,
$\Phi=(2m)^{-1/2}[e^{-imt}\psi_++e^{imt}\psi_-^*]$, the kinetic and potential energies are
$|\dot\Phi|^2=\tfrac m2[|\psi_+|^2+|\psi_-|^2-2\mathrm{Re}(e^{-2imt}\psi_+\psi_-)]$ and
$m^2|\Phi|^2=\tfrac m2[|\psi_+|^2+|\psi_-|^2+2\mathrm{Re}(e^{-2imt}\psi_+\psi_-)]$. Their oscillating parts cancel in the density and add in the
pressure:
\begin{equation}
  \rho=m\big(|\psi_+|^2+|\psi_-|^2\big),\qquad p=-2m\,\mathrm{Re}\big(e^{-2imt}\psi_+\psi_-\big).
  \label{eq:rhop_osc}
\end{equation}
The density has no component at $2m$, which justifies \eqref{eq:fastF}. The pressure averages to zero over a period, so the
pairs behave as non-relativistic matter on the slow scale. But it oscillates with the pair coherence $\psi_+\psi_-$. This is the
oscillating quantum pressure $\propto\mathrm{Re}(\alpha^*\beta e^{2imt})$ of the coherent pair state of
Refs.~\cite{ParkerFulling1973,Xue2023}. In a quantum state, $\psi_+\psi_-$ stands for the expectation value
$\langle\psi_+\psi_-\rangle=\int\frac{\dd^3k}{(2\pi)^3}c_k$.

\begin{figure}[t]
\centering
\includegraphics[width=\linewidth]{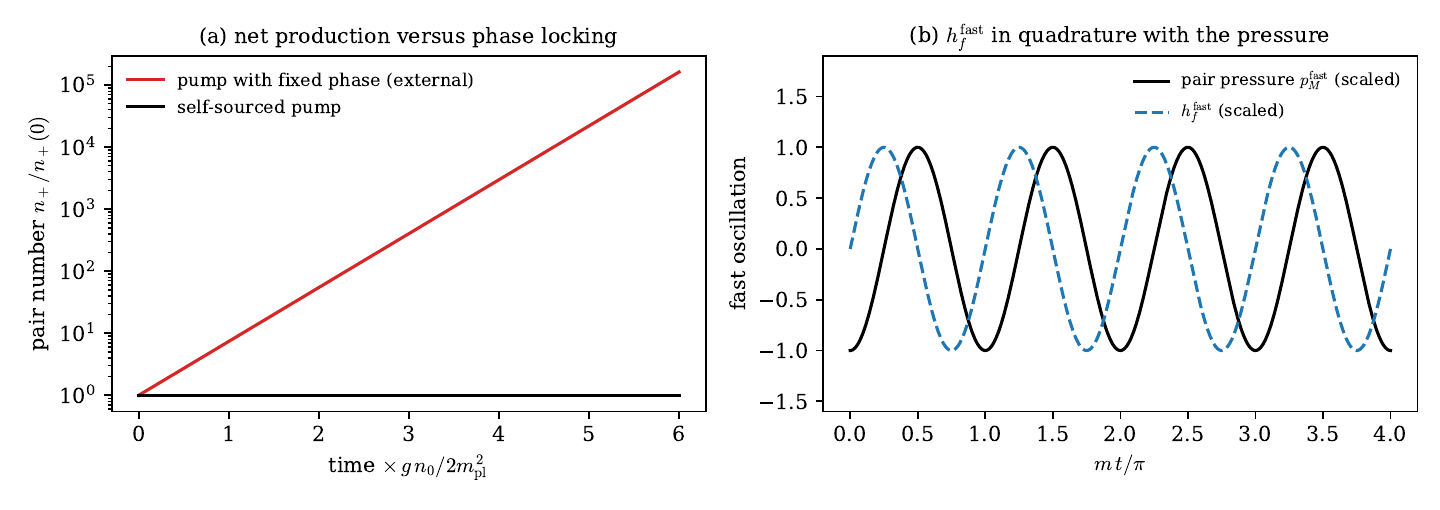}
\caption{(a)~Pair number for an external pump with fixed phase (exponential growth) and for the self-sourced pump of
Eq.~\eqref{eq:selfcons} (no net production). Both pumps have the same strength. Time is in units of $(g\,n_0/2\mpl^2)^{-1}$, the inverse
self-sourced pump strength for initial pair-coherence density $n_0$. (b)~Fast oscillations of the pair pressure and of $\hf{}$: the two are in quadrature,
which is the phase at which the pump exchanges no net energy with the pairs.}
\label{fig:locking}
\end{figure}

\paragraph{Self-consistent amplitude.} The oscillating pressure drives $\hf{}$ through \eqref{eq:fastF}. Writing
$\dot h_f^{\rm fast}\simeq\mathrm{Re}[-2im\tilde he^{-2imt}]$ and equating the coefficients of $e^{-2imt}$, we find
\begin{equation}
  \tilde h=\frac{i}{2\mpl^2}\,\psi_+\psi_- .
  \label{eq:selfcons}
\end{equation}
The oscillation of $H$ is in quadrature with the pressure, as $\dot h_f^{\rm fast}\propto-p$ requires (figure~\ref{fig:locking}b). Its magnitude is
fixed by the pairs. Let $n_c=|\psi_+\psi_-|$ be the pair-coherence density, and use $\rho_M=2mn$ for equal numbers of particles and
antiparticles. Then
\begin{equation}
  \eta=\frac{|\tilde h|}{\bar H}=\frac34\,\frac{\bar H}{m}\,\frac{\rho_M}{3\mpl^2\bar H^2}\,\frac{n_c}{n}\;\le\;\frac34\,\frac{\bar H}{m}\,\frac{n_c}{n}.
  \label{eq:eta}
\end{equation}
For a classical, fully coherent pair field $n_c=n$, and the relative amplitude is at most $\tfrac34\bar H/m$. This is consistent
with the relative oscillation amplitude $\sim10^{-3}$ found numerically in Ref.~\cite{Xue2023} for $\bar H/m=10^{-3}$. In a dilute
squeezed quantum state, $|c_k|=[f_k(1+f_k)]^{1/2}$ can exceed $f_k$, so $n_c/n$ can exceed one. Reaching $\eta=\Ocal(1)$ would, however,
require $n_c/n\sim m/\bar H$, together with phase alignment of all populated modes over a Hubble time, whereas the phases
$e^{-2i\delta_kt}$ disperse on the time $\sim m/p^2$.

\paragraph{Phase locking.} Substituting \eqref{eq:selfcons} into \eqref{eq:pump} at $k=0$ gives
$\dot\psi_\pm=(ig/2\mpl^2)|\psi_\mp|^2\psi_\pm$. The amplitudes $|\psi_\pm|$ are therefore constant, and only the phases rotate. The rate at
which the pump changes the pair number, $4g\,\mathrm{Re}(\tilde h\psi_+^*\psi_-^*)$, vanishes identically, because
$\tilde h\psi_+^*\psi_-^*=i|\psi_+\psi_-|^2/2\mpl^2$ is purely imaginary. The pair coherence, the pressure, $\hf{}$ and $\rho_\Lambda^{\rm fast}$ thus lock into a
stable collective oscillation, at the slightly shifted frequency $2m-g\rho_M/(2m\mpl^2)$. This is the non-decaying coherent
oscillation found numerically in Ref.~\cite{Xue2023}.

\paragraph{External versus self-sourced pump.} The contrast with an external pump explains why no pairs are produced. An
\emph{external} pump is an oscillation of $H$ imposed by something other than the pairs, for instance the oscillating inflaton of
standard reheating (section~\ref{sec:reheating_intro}). Its amplitude and phase are fixed independently of the pairs and are not affected by
them. In \eqref{eq:pump} the growing solution then automatically selects the relative phase between $\tilde h$ and $\psi_+\psi_-$ that
maximises the pumping, and the pair number grows as $e^{2\mu_0t}$. The energy comes from the external source, which is not depleted
in this approximation. A \emph{self-sourced} pump is the oscillation of $H$ generated by the pressure of the same pairs it acts on,
through \eqref{eq:selfcons}. Its phase is then fixed by the pair coherence, at $90^\circ$ from it, which is precisely the phase at
which the pumping vanishes. The pairs cannot pump themselves, because the closed pair--gravity system has no external energy
source on the fast time scale.

Figure~\ref{fig:locking}a shows the two cases with pumps of the \emph{same} strength. Time is measured in units of
$(g\,n_0/2\mpl^2)^{-1}$, which is the inverse of the self-sourced pump strength $g|\tilde h|$ for an initial pair-coherence density $n_0$.
The external pump has the same strength, $\Ups=g|\tilde h|$, but a fixed phase. Starting from the same pair state, the external pump
increases the pair number by five orders of magnitude within six time units, while the self-sourced pump leaves it constant to
machine precision. Figure~\ref{fig:locking}b shows the origin of the difference: the fast oscillations of the pressure and of
$\hf{}$ are in quadrature.

\paragraph{Consequences.} A self-sourced oscillation of $H$ exchanges no net energy with the pairs: it settles at exactly
the phase for which the pumping vanishes. Net conversion of the energy of the cosmological term into pairs requires a drive whose phase is not
locked to the pairs. On the slow scale this drive is the decrease of $\rho_\Lambda$, which section~\ref{sec:coherent} treats. Moreover,
the bound \eqref{eq:eta} limits the resonant mechanism: with a self-sourced oscillation the production rate
\eqref{eq:Gprod} is at most $\tfrac98\bar H^2/m$ (for $\xi=0$ and $n_c=n$), far below the Hubble rate.

\section{Macroscopic equations from the fast coherent dynamics}\label{sec:coherent}

In this section we keep only the fast coherent oscillation. The fluctuation in \eqref{eq:La} is the deterministic function
$\delta\Ocal_S=(6\xi-\tfrac32)\dot h_f^{\rm fast}$, and there is no stochastic fluctuation of the spacetime vacuum. We write the complete microscopic system, solve it by a
two-timescale expansion, discuss the meaning of the solution, and average it over the fast time.

\subsection{The complete microscopic system}

The system consists of five parts:
\begin{enumerate}[label=(\alph*),leftmargin=1.6em,itemsep=2pt]
  \item the metric, $H=\bar H+\hf{}$;
  \item the kinetic equations \eqref{eq:kin} for every mode, with $\Omega_k$ evaluated with the full $H$, so that $\hf{}$ enters through
  $\Delta$ and through $\dot\Omega_k$;
  \item the density and pressure \eqref{eq:rhop} of the pairs;
  \item the exact Einstein equations \eqref{eq:F1}--\eqref{eq:F2}, with $\rho_\Lambda$ fixed by the energy balance \eqref{eq:bianchi};
  \item the decays, which add $-\Gde f_k$ to $\dot f_k$, damp the coherence as $-\tfrac12\Gde c_k$, and feed the radiation at the rate $\Gde\rho_M$.
\end{enumerate}
The fast part of the Einstein equations closes the system. In terms of the mode coherences, \eqref{eq:fastF} reads
\begin{equation}
  \dot h_f^{\rm fast}=\frac{g_\Phi}{\mpl^2a^3}\int\!\frac{\dd^3k}{(2\pi)^3}\,\Omega_k\,\mathrm{Re}\,c_k^{\rm osc},\qquad \rho_\Lambda^{\rm fast}=6\mpl^2\bar H\hf{} ,
  \label{eq:closure}
\end{equation}
where $c_k^{\rm osc}$ is the part of the coherence that oscillates at $2\Omega_k$ (defined below). For a single coherent mode this reduces to
\eqref{eq:selfcons}.

\subsection{Two-timescale solution}\label{sec:split}

The kinetic equations \eqref{eq:kin} are linear and inhomogeneous in the three real variables $f_k$, $\mathrm{Re}\,c_k$ and $\mathrm{Im}\,c_k$.
The inhomogeneous term $\dot\Omega_k/2\Omega_k$ comes from the ``1'' in $1+2f_k$, i.e.\ from the spacetime vacuum. For $\zeta_k\ll1$ the general
solution is the sum of a particular solution and three homogeneous ones, of which one varies slowly and two rotate at
$2\Omega_k$. Correspondingly (appendix~\ref{app:twotime}),
\begin{equation}
  f_k=f_k^{\rm ad}+f_k^{\rm r}+f_k^{\rm osc},\qquad c_k=c_k^{\rm ad}+c_k^{\rm osc}.
  \label{eq:split}
\end{equation}
The three pieces have a clear physical meaning.
\begin{itemize}[leftmargin=1.5em,itemsep=3pt]
  \item The \emph{adiabatic part} is the particular solution driven by the spacetime vacuum. It is the occupation of
  section~\ref{sec:rhoH}, enhanced by Bose stimulation, $f_k^{\rm ad}=(\dot\Omega_k/4\Omega_k^2)^2(1+2f^{\rm r}_k)$. It follows the instantaneous
  background with no delay, and its energy density is $\rhoMH=2\chi m^2\bar H^2$.
  \item The \emph{real pairs} $f_k^{\rm r}$ are the slow homogeneous solution. They are an adiabatic invariant: to every order in
  $\bar H/m$ the slow change of $f_k$ is a total time derivative of a local function of the background, so $f_k^{\rm r}$ is constant
  and the comoving number of real pairs is conserved. Physically, with $E_k\equiv\langle|\dot\chi_k|^2\rangle+\Omega_k^2\langle|\chi_k|^2\rangle=\Omega_k(1+2f_k)$ the energy of the mode pair $(\kk,-\kk)$, including its
zero-point energy $\Omega_k$, the quantity $f_k=E_k/2\Omega_k-\tfrac12$ counts the quanta of an oscillator
  whose frequency changes slowly, and that number does not change. The real pairs define the number and energy densities
  \begin{equation*}
    n^{\rm r}=\frac{2g_\Phi}{a^3}\int\!\frac{\dd^3k}{(2\pi)^3}\,f^{\rm r}_k,\qquad
    \rho^{\rm r}=\frac{2g_\Phi}{a^3}\int\!\frac{\dd^3k}{(2\pi)^3}\,\Omega_k\,f^{\rm r}_k ,
  \end{equation*}
  i.e.\ the part of the pair density that is \emph{not} the adiabatic dressing $\rhoMH$. Its evolution is given in Eq.~\eqref{eq:real}.
  \item The \emph{oscillating part} consists of the two fast homogeneous solutions,
  $c^{\rm osc}_k=|c^0_k|e^{-2i\theta_k+i\varphi_k}$, with $f_k^{\rm osc}=(\dot\Omega_k/2\Omega_k^2)|c^0_k|\sin(2\theta_k-\varphi_k)$. It is the coherent
  back-and-forth creation and annihilation of pairs at frequency $2\Omega_k$. Its average vanishes mode by mode, which is the
  detailed balance of Ref.~\cite{Xue2023b}.
\end{itemize}
The resonant exchange with $\hf{}$ averages to zero when $\hf{}$ is self-sourced (section~\ref{sec:backreact}).

The three pieces are not independent: they are tied together by the invariant $\mathcal P_k=|c_k|^2-f_k(1+f_k)$, which
\eqref{eq:kin} conserves exactly. For a state evolved from the spacetime vacuum $\mathcal P_k=0$: the state is pure, a two-mode squeezed
state of the particle in $\kk$ and the antiparticle in $-\kk$. Since $c^{\rm ad}$ and $f^{\rm ad}$ are small, of order $\zeta_k$ and $\zeta_k^2$, to leading
order $\mathcal P_k\simeq|c^{\rm osc}_k|^2-f^{\rm r}_k(1+f^{\rm r}_k)$. Hence $|c^{\rm osc}_k|^2=f^{\rm r}_k(1+f^{\rm r}_k)$: a non-adiabatic event that creates real pairs
necessarily creates, at the same time, a phase correlation between each particle and its antiparticle. This coherence makes
the pair number oscillate back and forth at $2\Omega_k$, and its sum over modes is the oscillating pressure that sources
$\hf{}$. Real pairs \emph{without} coherence ($|c^{\rm osc}|^2<f^{\rm r}(1+f^{\rm r})$, i.e.\ $\mathcal P_k<0$) describe a mixed state. On macroscopic
scales such a state arises effectively when the phases of many modes disperse: the summed coherence is lost by phase
mixing, while each $f^{\rm r}_k$ survives (section~\ref{sec:dressing}).

\subsection{Adiabatic dressing, real pairs and numerical tests}\label{sec:dressing}

The decomposition \eqref{eq:split} separates two things that are easily confused: the power-law density $\rhoMH\propto m^2H^2$
and the exponentially suppressed creation of real pairs. Both come from the same Bogoliubov coefficient. Integrating
$\beta_k=\int^t(\dot\Omega_k/2\Omega_k)e^{-2i\theta_k}$ by parts repeatedly produces a series of boundary terms at the upper limit $t$,
$i\dot\Omega_k/4\Omega_k^2+\Ocal(\zeta_k^2)+\dots$, plus a remainder $R_k$. The boundary terms depend only on the background at the same
instant. Their square is the adiabatic occupation, and they make up $\rhoMH$. The series is asymptotic; truncated at its
smallest term, it leaves a remainder controlled by the singularities of $\Omega_k(t)$ in complex time, $|R_k|^2\sim e^{-c\,\Omega_k/H}$
with $c\simeq2\pi$. When the expansion stops, the boundary terms vanish and only $|R_k|^2$ survives. The dressing is reversible,
and the real pairs are exponentially few. In a higher-order adiabatic basis the dressing is absorbed into the definition of
the spacetime vacuum, and in the optimal (superadiabatic) basis the occupation is exponentially small at all times and rises smoothly
to $|R_k|^2$~\cite{Berry1989,DabrowskiDunne2014,DabrowskiDunne2016}. Only $|R_k|^2$ is independent of the basis.

\begin{figure}[t]
\centering
\includegraphics[width=\linewidth]{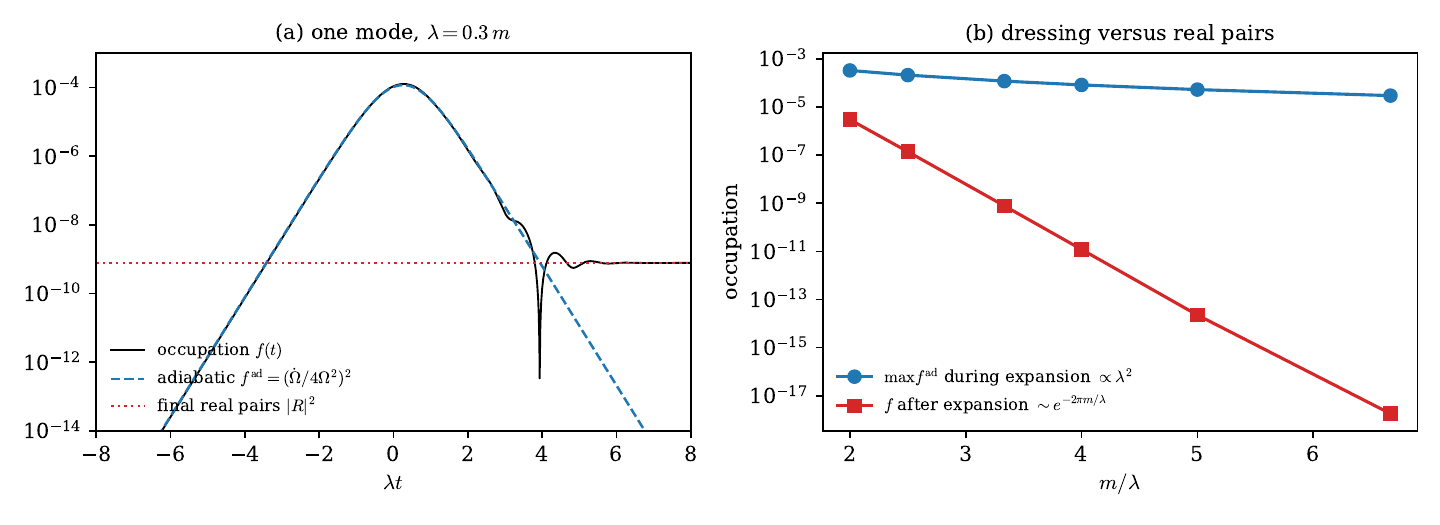}
\caption{Adiabatic dressing versus real pairs, for the mode \eqref{eq:tanhmodel} whose momentum decreases from $p_0=m$ to $0$ during a
time $\sim\lambda^{-1}$.
(a)~While the frequency changes, the occupation follows the adiabatic value $f^{\rm ad}$. When the expansion stops, the dressing
disappears and only the real pairs created in the smoothed Stokes jump remain. (b)~The peak dressing scales as $\lambda^2$, a power
law (the analogue of $\rhoMH\propto m^2H^2$); the real pairs scale as $e^{-2\pi m/\lambda}$ (fitted slope $-6.06$).}
\label{fig:dressing}
\end{figure}

Figure~\ref{fig:dressing} illustrates this with a single mode whose frequency changes once, smoothly, between two constant
values:
\begin{equation}
  \Omega^2(t)=m^2+\tfrac12p_0^2\big[1-\tanh(\lambda t)\big].
  \label{eq:tanhmodel}
\end{equation}
The physical momentum decreases from $p_0$ at $t\to-\infty$ to $0$ at $t\to+\infty$, as for a momentum redshifted by an expansion that
starts and then stops. The parameter $\lambda$ is the \emph{rate} of this change: the momentum changes during a time $\sim\lambda^{-1}$ around
$t=0$, and $|\dot\Omega/\Omega|\sim\lambda\,p_0^2/\Omega^2$. In an expanding universe the corresponding rate is $|\dot\Omega_k/\Omega_k|=Hp^2/\Omega_k^2$, so $\lambda$ plays the
role of the Hubble rate $H$. Unlike an expanding universe, however, \eqref{eq:tanhmodel} has a definite final state, in which the
number of real pairs is unambiguous. We take $p_0=m$. During the change, the exact solution of \eqref{eq:kin} follows $f^{\rm ad}=(\dot\Omega/4\Omega^2)^2$, whose peak
scales as $\lambda^2$ (fitted exponent $2.000$). After the change, the occupation settles to a constant that scales as
$e^{-6.1\,m/\lambda}$, close to $e^{-2\pi m/\lambda}$. The same mode thus carries a power-law dressing while the background changes, and an
exponentially small number of real pairs once it stops.

Summed over all modes, the dressing is a term of second adiabatic order in $\langle T_{\mu\nu}\rangle$. In a covariant regularisation it
is proportional to the Einstein tensor and is absorbed into Newton's constant~\cite{ParkerFulling1974,AndersonParker1987,BirrellDavies}.
What then remains for $m\gg H$ is a local curvature correction suppressed as $H^6/m^2$~\cite{AppelquistCarazzone1975}, plus the
exponentially small real part. The finite $\rhoMH=2\chi m^2H^2$ arises from the non-relativistic cut-off. Reading it as a stationary
population of \emph{real} pairs, as the scenario of Refs.~\cite{Xue2023b,Xue2025} requires, needs production that is not
exponentially suppressed. In the present framework this is available through a drive with spectral weight at $2m$
(section~\ref{sec:resonance}), through fluctuations of the spacetime vacuum far from the de Sitter state (section~\ref{sec:kinetic} and
appendix~\ref{app:kinetic}), or through dissipation (section~\ref{sec:averaged}).

\begin{figure}[t]
\centering
\includegraphics[width=\linewidth]{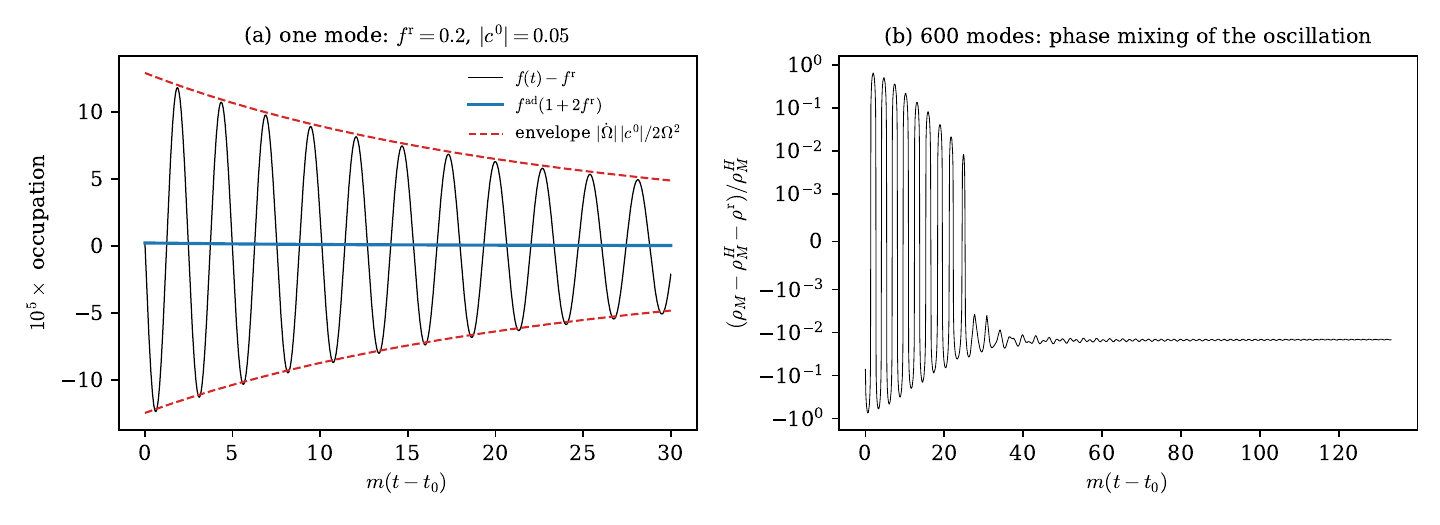}
\caption{(a)~A single mode with real pairs $f^{\rm r}=0.2$ and initial coherence $|c^0|=0.05$: the occupation oscillates around the
adiabatic value $f^{\rm ad}(1+2f^{\rm r})$ within the predicted envelope $|\dot\Omega||c^0|/2\Omega^2$, while $f^{\rm r}$ stays constant. (b)~For 600
modes in a matter-dominated background, the oscillation of the total density around $\rhoMH+\rho^{\rm r}$ decays by phase mixing. The
small constant offset of $1.5\%$ comes from the first-order accuracy of the initial state.}
\label{fig:modes}
\end{figure}

Figure~\ref{fig:modes} tests the decomposition \eqref{eq:split} directly. In both panels, what is \emph{calculated} is the exact
numerical solution of the kinetic equations \eqref{eq:kin}. What is \emph{predicted} are the analytic results of
section~\ref{sec:split}: the adiabatic occupation $f^{\rm ad}=(\dot\Omega/4\Omega^2)^2(1+2f^{\rm r})$, the conservation of $f^{\rm r}$, the oscillation
envelope $|\dot\Omega||c^0|/2\Omega^2$, and the coefficient of $\rhoMH$.

\emph{Panel (a): a single mode.} The frequency is $\Omega^2=m^2+p_0^2(t_0/t)^{4/3}$: the momentum redshifts as $a^{-1}\propto t^{-2/3}$, as in a
matter-dominated universe, with $p_0=0.8m$ at $t_0=40\,m^{-1}$. The initial state contains real pairs, $f^{\rm r}=0.2$, and an initial
coherence $|c^0|=0.05$, on top of the adiabatic part. We integrate \eqref{eq:kin} and plot $f(t)-f^{\rm r}$. The calculated occupation
oscillates at $2\Omega$ around the predicted adiabatic value and stays within the predicted envelope. The calculated amplitude,
$2.9\times10^{-5}$ averaged over the late window, agrees with the predicted $2.7\times10^{-5}$ to within $8\%$; the difference comes from
higher-order terms and from averaging the slowly varying envelope. In separate runs, a real population $f^{\rm r}=0.5$ is
conserved to $10^{-7}$, and the invariant $|c|^2-f(1+f)$ to $10^{-14}$.

\emph{Panel (b): an ensemble of modes.} We integrate \eqref{eq:kin} for 600 comoving modes in a matter-dominated background, with
$\bar H$ decreasing from $0.02\,m$ to $0.004\,m$. The initial state is the adiabatic one, plus real pairs and an initial coherence
concentrated around $p=0.5\,m$ (amplitudes $10^{-4}$ and $3\times10^{-3}$). From the calculated occupations we compute $\rho_M$ with
\eqref{eq:rhop}, and compare it with the predicted $\rhoMH$ (from $f^{\rm ad}$) plus the energy $\rho^{\rm r}$ of the real pairs. The panel
shows $(\rho_M-\rhoMH-\rho^{\rm r})/\rhoMH$. The fast oscillation, initially of order $\rhoMH$, decays by phase mixing to the $10^{-3}$ level,
because different modes oscillate at different frequencies $2\Omega_k$. After the transient the calculated density agrees with the
prediction to $0.13\%$. The comoving number of real pairs is conserved to $5\times10^{-4}$. The calculated ratio $\rhoMH/(g_\Phi m^2\bar H^2)$
stays between $2.25\times10^{-4}$ and $2.28\times10^{-4}$, against the predicted $I(1)/16\pi^2=2.26\times10^{-4}$.

\subsection{Averaged equations and the stationary occupation}\label{sec:averaged}

We now average the microscopic system over the fast period (appendix~\ref{app:twotime}).

\paragraph{Friedmann equations.} Averaging $H^2=(\bar H+\hf{})^2$ gives the first Friedmann equation for the averaged quantities,
corrected only by $-\langle(\hf)^2\rangle$, whose relative size $\eta^2/2\lesssim\tfrac{9}{32}(\bar H/m)^2$ is about $10^{-8}$. Averaging the $\dot H$
equation removes every term linear in fast quantities, and the averaged pressure is the kinetic pressure of the pairs,
$w_M=\langle p_M\rangle/\bar\rho_M\ll1$. Equations \eqref{eq:F1}--\eqref{eq:F2} therefore hold for the averaged quantities.

\paragraph{Energy balance.} The average of the Bianchi identity contains the correlator $3\langle \hf{}(\rho_M^{\rm fast}+p_M^{\rm fast})\rangle$. It
vanishes because $\rho_M^{\rm fast}$ has no component at $2m$ and $\hf{}$ is in quadrature with $p_M^{\rm fast}$. Equation \eqref{eq:bianchi}
therefore also holds for the averaged quantities.

\paragraph{Pair equation.} The exact energy balance of the pairs is
$\dd(a^3\rho_M)/\dd t=2g_\Phi\int\frac{\dd^3k}{(2\pi)^3}\dot\Omega_k(f_k+\mathrm{Re}\,c_k)$: the first term is the change of energy of existing quanta
as their frequency changes, the second the creation or annihilation of quanta. Inserting \eqref{eq:split} and averaging,
three things happen. The adiabatic terms combine into $\dd(a^3\rhoMH)/\dd t$. The real pairs do only the work of their kinetic
pressure. The oscillating and resonant terms average to zero. Since $\rhoMH\propto\bar H^2$ and $\dot{\bar H}=-\epsilon\bar H^2$, the power
needed to keep the density on $\rhoMH$ is $(\partial_t+3\bar H)\rhoMH=(3-2\epsilon)\bar H\rhoMH$.

The decays add one more effect. Damping the coherence shifts the phase of the slaved part, giving
$c_k\simeq(\dot\Omega_k/2\Omega_k)(1+2f_k)/(2i\Omega_k+\tfrac12\Gde)$. To first order in $\Gde/\Omega_k$ this has a real part, and the spacetime-vacuum source then
creates quanta at the rate $\Gde(\dot\Omega_k/4\Omega_k^2)^2(1+2f_k)=\Gde f^{\rm ad}_k$. This is exactly the rate at which they decay. The
adiabatic population is therefore not depleted by the decays. Instead, energy flows from the background through it into
radiation at the rate $\Gde\rhoMH$. Numerically, with $\Gde=0.01$, $0.03$ and $0.1\,m$, we find $\langle f\rangle/f^{\rm ad}=1.016$, $1.0007$
and $0.9996$. Altogether,
\begin{equation}
  \dot{\bar\rho}_M+3\bar H(1+w_M)\bar\rho_M=\mathcal Q-\Gde\,\bar\rho_M,\qquad
  \mathcal Q=\big[(3-2\epsilon)\bar H+\Gde\big]\rhoMH=-\dot{\bar\rho}_\Lambda,
  \label{eq:rate_coh}
\end{equation}
with $\bar\rho_M=\rhoMH+\rho^{\rm r}$. Here $\rhoMH$ is the adiabatic dressing, which follows $\bar H$ instantaneously, and $\rho^{\rm r}$ is the
energy density of the real pairs defined in section~\ref{sec:split}. Because $f^{\rm r}_k$ is an adiabatic invariant, the real pairs neither
follow $\rhoMH$ nor relax towards it. Their density obeys
\begin{equation}
  \dot\rho^{\rm r}+3\bar H\,(1+w^{\rm r})\,\rho^{\rm r}=-\Gde\,\rho^{\rm r},\qquad w^{\rm r}=\frac{\int\dd^3k\,(p^2/3\Omega_k)f^{\rm r}_k}{\int\dd^3k\,\Omega_kf^{\rm r}_k}\ll1 .
  \label{eq:real}
\end{equation}
Without decays, the comoving number $a^3n^{\rm r}$ is exactly conserved, and $\rho^{\rm r}\propto a^{-3(1+w^{\rm r})}\simeq a^{-3}$: the real pairs
evolve as non-relativistic matter, losing energy only through the work of their small kinetic pressure. The decays deplete them
as $e^{-\Gde t}$. This is in sharp contrast with the rate equation \eqref{eq:rate}, in which any excess over $\rhoMH$ relaxes at the rate
$\GaM$. Real massive pairs of this kind are produced in the reheating epoch. There the expansion departs from detailed balance
($\GaM$ is no longer much larger than $H$) and from adiabaticity, so that real pairs $f^{\rm r}_k$ are created. Once created, they evolve
according to \eqref{eq:real}: as non-relativistic matter, decaying into radiation~\cite{Xue2023b}. The radiation equation is unchanged.

\paragraph{Stationary occupation.} In the adiabatic case (section~\ref{sec:kinetic}), $\rhoMH$ is the energy of the stationary occupation
$f^*_k=[(\Gamma^\downarrow_k/\Gamma^\uparrow_k)^{1/2}-1]^{-1}$ of the kinetic equation. That stationary state exists because the fluctuations of the spacetime
vacuum absorb more readily than they emit, $S_{\mathcal O}(+2m)>S_{\mathcal O}(-2m)$, i.e.\ $\Gamma^\downarrow>\Gamma^\uparrow$. In the fast-oscillating case, the source
$\delta\Ocal_S=(6\xi-\tfrac32)\dot h_f^{\rm fast}$ is a deterministic, classical function. The spectrum of a classical function is symmetric,
$S(\omega)=S(-\omega)$, so $\Gamma^\uparrow=\Gamma^\downarrow$ and the expression for $f^*$ diverges: no finite stationary occupation exists. In the coherent regime, modes
in the resonance band grow until back-reaction locks the phases, and the locked state then keeps \emph{any} occupation constant.

What replaces $f^*$ is the adiabatic occupation, $f^{\rm ad}_k\simeq H^2p^4/16\Omega_k^6$. It is not a stationary state of a kinetic equation but
the instantaneous dressing of the spacetime vacuum, and by construction it reproduces $\rhoMH$. For $H/m=1.47\times10^{-4}$ it equals
$H^2/128m^2=1.7\times10^{-10}$ at $p=m$, and the uniform occupation of the non-relativistic shell with the same energy is
$1.15\times10^{-10}$. Inserting $f^{\rm ad}_k$ into the energy integral gives $\rhoMH=2.263\times10^{-4}g_\Phi m^2H^2$, equal to $g_\Phi I(1)m^2H^2/16\pi^2$.

\emph{Does the fast oscillation change this occupation?} The occupation $f^{\rm ad}_k$ is the dressing produced by the \emph{slow}
change of $\Omega_k$ through the expansion. The fast oscillation $\hf{}$ produces an additional occupation of its own, given by the
pump solution of section~\ref{sec:resonance}. Outside the resonance band ($p>p_{\rm res}$) this occupation oscillates with time average
$\Ups^2/2\nu_k^2$; inside the band it grows as $\sinh^2(\mu_kt)$. To compare the two, we proceed in three steps.
\begin{enumerate}[label=(\roman*),leftmargin=1.6em,itemsep=2pt]
  \item For every momentum $p\le m$ we take the exact occupation $n_k(t)$ produced from the spacetime vacuum by a constant pump of
  strength $\Ups=|g|\eta\bar H$ [Eq.~\eqref{eq:nk} and its continuation outside the band], with $\xi=0$ and $H/m=1.47\times10^{-4}$.
  \item We average $n_k(t)$ over one Hubble time, $0\le t\le\bar H^{-1}$, which is the longest time over which the pump can act
  before the background changes.
  \item We weight the averaged occupation with the mode energy, integrate over the shell as in \eqref{eq:rhop}, and divide by $\rhoMH$.
\end{enumerate}
The result depends only on the relative amplitude $\eta$:
\begin{center}\small
\begin{tabular}{@{}lccc@{}}
\toprule
 & $\eta$ & $\Ups/\bar H=|g|\eta$ & pump-induced energy$/\rhoMH$\\
\midrule
self-sourced, Eq.~\eqref{eq:eta} & $1.1\times10^{-4}$ & $8.3\times10^{-5}$ & $4.2\times10^{-4}$\\
external drive & $10^{-2}$ & $7.5\times10^{-3}$ & $3.4$\\
external drive & $1$ & $0.75$ & $3.7\times10^{4}$\\
\bottomrule
\end{tabular}
\end{center}
For the self-sourced oscillation the extra occupation is less than $10^{-3}$ of the adiabatic one, and no mode grows appreciably
within a Hubble time because $\Ups/\bar H\sim10^{-4}$. The occupation that gives $\rhoMH$ is therefore $f^{\rm ad}_k$, unaffected by the fast
oscillation. Only an oscillation driven from outside with $\eta\gtrsim10^{-2}$ would dominate. Its occupation would then scale as
$\eta^2$ and be concentrated near the band edge $p\sim p_{\rm res}$, rather than follow $m^2H^2$; that is a different physical situation, not
the one considered here.

\section{The closed set of background equations (fast-oscillating case)}\label{sec:closed2}

Collecting the averaged equations, the closed set of background equations in the fast-oscillating case is
\begin{equation}
  \boxed{
  \begin{gathered}
  \bar H^2=\frac{\bar\rho_\Lambda+\bar\rho_M+\bar\rho_R}{3\mpl^2},\qquad
  \dot{\bar H}=-\frac{\bar\rho_M+\frac43\bar\rho_R}{2\mpl^2},\\[2pt]
  \dot{\bar\rho}_M+3\bar H\bar\rho_M=\big[(3-2\epsilon)\bar H+\Gde\big]2\chi m^2\bar H^2-\Gde\bar\rho_M,\qquad
  \dot{\bar\rho}_R+4\bar H\bar\rho_R=\Gde\bar\rho_M ,
  \end{gathered}}
  \label{eq:closed2}
\end{equation}
with $\bar\rho_M=2\chi m^2\bar H^2+\rho^{\rm r}$, the real pairs evolving according to \eqref{eq:real}, and $-\dot{\bar\rho}_\Lambda=[(3-2\epsilon)\bar H+\Gde]2\chi m^2\bar H^2$. As in Part~I, the four equations are
independent and determine the four unknowns $(\bar H,\bar\rho_\Lambda,\bar\rho_M,\bar\rho_R)$; the energy transfer $-\dot{\bar\rho}_\Lambda$ follows from them
through the Bianchi identity. The difference lies in the source of the pair equation. The relaxation term $\GaM(\rhoMH-\rho_M)$ of
\eqref{eq:closed1} is replaced by an explicit source fixed by the background: the power needed to hold the pairs on $\rhoMH$, plus
the power that flows through them into radiation.

\paragraph{Exact solution.} Without decays and real pairs, \eqref{eq:closed2} has the exact solution \eqref{eq:exact} found in
section~\ref{sec:exact}, with constant $\epsilon=\chi m^2/\mpl^2$. In the adiabatic case that solution required the detailed-balance
regime $\GaM\gg H$. In the fast-oscillating case it follows directly from the instantaneous tracking of $\rhoMH$.

\paragraph{Relation to the cosmic rate equation.} To compare \eqref{eq:closed2} with \eqref{eq:closed1}, we return to the detailed-balance
discussion of section~\ref{sec:exact} and Eq.~\eqref{eq:detbal}, and make it systematic. Write $\rho_M=\rhoMH+\delta$ in the rate
equation \eqref{eq:rate}. It becomes $\dot\delta+\Gtot{}\delta=-S$, where $S=[(3-2\epsilon)H+\Gde]\rhoMH$ is the power needed to hold the pairs on
$\rhoMH$ and $\Gtot{}=3H+\GaM+\Gde$ is the total rate at which a deviation disappears. The solution has two parts:
\begin{itemize}[leftmargin=1.5em,itemsep=2pt]
  \item a \emph{slaved} part, $\delta_{\rm s}=-S/\Gtot{}+\Ocal(H/\Gtot^2)$, which follows the source with a lag of order $1/\GaM$;
  \item a \emph{free} part, which carries any initial excess of pairs and decays at the rate $\Gtot{}$.
\end{itemize}
The coherent dynamics has no lag, $\delta_{\rm s}=0$. Its free part, the real pairs, decays only at $3H+\Gde$. The two descriptions
therefore coincide when $\GaM\gg H,\,\Gde$, with $\GaM(\rhoMH-\rho_M)$ finite, and no initial excess. This is precisely the detailed-balance regime \eqref{eq:detbal}: the first-order lag in
\eqref{eq:detbal} is $\delta_{\rm s}=-S/\GaM$, and it is negligible for $\GaM\gg H$. In this regime a fast relaxation acting on a small
deviation carries the finite energy flux $\GaM(\rhoMH-\rho_M)\to S$. At finite $\GaM$ the source of
the rate equation is $S[1-(3H+\Gde)/\GaM+\Ocal(H/\GaM)]$.

\begin{figure}[t]
\centering
\includegraphics[width=\linewidth]{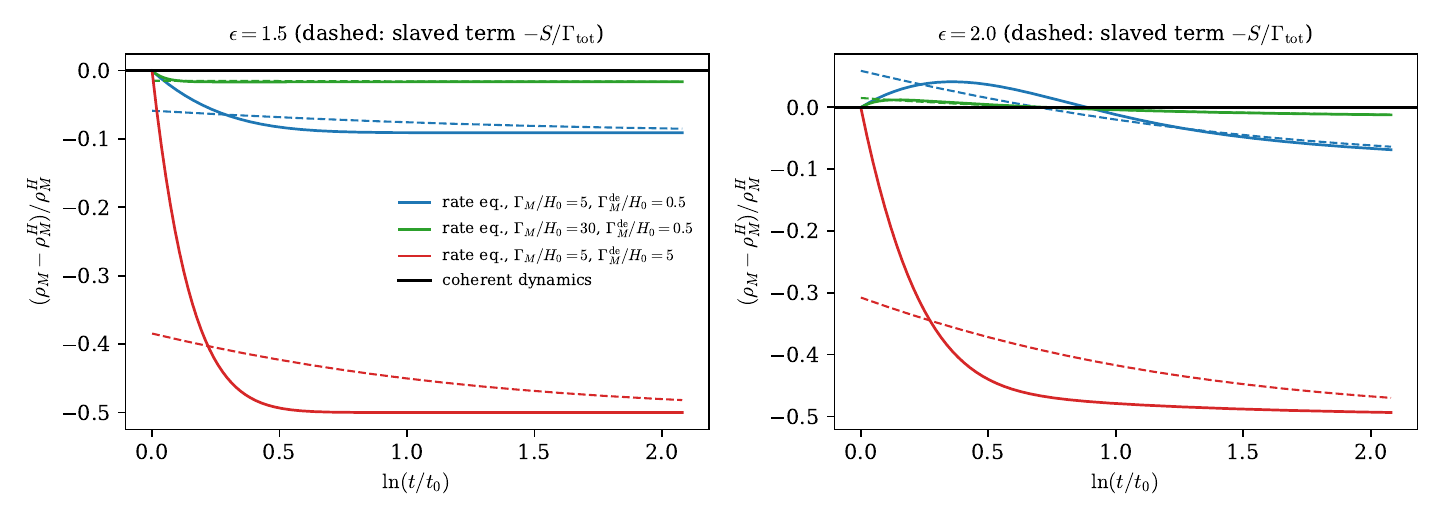}
\caption{Deviation of the pair density from $\rhoMH$ in the cosmic rate equation (solid), for constant $\GaM$ and $\Gde$ and two values of
$\epsilon$. Dashed lines show the slaved term $-S/\Gtot{}$; the coherent dynamics gives exactly zero (thick black line). The rate equation
approaches the coherent result as $\GaM/H$ grows, and departs from it at order one when $\Gde$ is comparable to $\GaM$.}
\label{fig:compare}
\end{figure}

Figure~\ref{fig:compare} and the following table illustrate this. The table gives the deviation at the final time, and the
rate equation reproduces the slaved term to $\Ocal(H/\Gtot{})$:
\begin{center}\small
\begin{tabular}{@{}cccccc@{}}
\toprule
$\epsilon$ & $\GaM/H$ & $\Gde/H$ & $(\rho_M-\rhoMH)/\rhoMH$ (rate eq.) & $-S/\Gtot{}\rhoMH$ & coherent\\
\midrule
1.5 & 60 & 6 & $-0.091$ & $-0.087$ & 0\\
1.5 & 600 & 6 & $-0.0099$ & $-0.0099$ & 0\\
2 & 60 & 6 & $-0.077$ & $-0.073$ & 0\\
2 & 600 & 6 & $-0.0083$ & $-0.0082$ & 0\\
2 & 12 & 30 & $-0.71$ & $-0.64$ & 0\\
\bottomrule
\end{tabular}
\end{center}
In the pair-dominated episode ($\epsilon=\tfrac32$, $\GaM\gg H\gg\Gde$) the two descriptions agree, with $\mathcal Q\simeq\Gde\rhoMH$ in both. When the
decays are fast compared with $\GaM$, the rate equation depletes the pairs, $\rho_M\ll\rhoMH$, whereas the coherent dynamics keeps
$\rho_M=\rhoMH$ and feeds the decays. The closed set \eqref{eq:closed2} is therefore the detailed-balance regime of \eqref{eq:closed1}. The
finite relaxation rate $\GaM$, and the finite lag it implies, are not generated by the coherent dynamics. They require
irreversible exchange with fluctuations of the spacetime vacuum that have a continuous spectrum, as in section~\ref{sec:kinetic}. This is the
Markov limit in which a reversible Hamiltonian evolution is replaced by an irreversible kinetic one~\cite{KlugerMottolaEisenberg1998}.

\section{Summary}\label{sec:summary}

We have derived the closed set of background equations for a Friedmann universe with a time-varying cosmological term and
massive particle--antiparticle pairs, starting from a complex scalar field with a global U(1) symmetry and a non-minimal
coupling to curvature. Table~\ref{tab:summary} compares the two cases.

\begin{table}[h]
\centering\small
\begin{tabular}{@{}p{0.2\linewidth}p{0.36\linewidth}p{0.36\linewidth}@{}}
\toprule
 & adiabatic case (Part I) & fast-oscillating case (Part II)\\
\midrule
Friedmann eqs. & from $G_{\mu\nu}+\Lambda(t)g_{\mu\nu}=8\pi GT_{\mu\nu}$ & the same, for averaged quantities, up to $\Ocal(\bar H^2/m^2)$\\[2pt]
energy balance & Bianchi identity & the same; $\langle \hf{}(\rho^{\rm fast}+p^{\rm fast})\rangle=0$\\[2pt]
pair density $\rhoMH$ & $2\chi m^2H^2$, $\chi=g_\Phi I(\kappa)/32\pi^2$ & the same, tracked instantaneously\\[2pt]
source term & $\GaM(\rhoMH-\rho_M)$, $\GaM=2(\Gamma^\uparrow\Gamma^\downarrow)^{1/2}$ & $[(3-2\epsilon)H+\Gde]\rhoMH$\\[2pt]
decay & $\Gde=y^2m/8\pi$, U(1)-invariant Yukawa & the same; adiabatic pairs replenished\\[2pt]
closed set & Eq.~\eqref{eq:closed1}; exact solution $\epsilon=\chi m^2/\mpl^2$ for $\GaM\gg H$ & Eq.~\eqref{eq:closed2}; the same exact solution\\
\bottomrule
\end{tabular}
\caption{The two closed sets of background equations.}
\label{tab:summary}
\end{table}

In the adiabatic case, the pairwise coupling of the rescaled field to the spacetime vacuum, the adiabatic occupation, a kinetic
equation for the coupling to fluctuations of the spacetime vacuum, and a U(1)-invariant Yukawa decay together give the cosmic rate equation and the
closed set \eqref{eq:closed1}. The relaxation rate is identified as $\GaM=2(\Gamma^\uparrow\Gamma^\downarrow)^{1/2}$ for dilute pairs. It is
proportional to $m$ if the fluctuations of the spacetime vacuum are scale-free at the pair-creation frequency. In the
detailed-balance regime, the closed set has an exact power-law solution with $\epsilon=\chi m^2/\mpl^2$.

In the fast-oscillating case, the oscillation of the Hubble function drives resonant pair production with coupling
$g=\tfrac12(\tfrac32-6\xi)$ and per-pair rate $|\tfrac32-6\xi|\eta\bar H$, without $e^{-m/H}$ suppression. A self-sourced oscillation, however, has
$\eta\lesssim\tfrac34\bar H/m$ and locks in phase with the pairs, exchanging no net energy with them. Averaging the coherent dynamics
gives the closed set \eqref{eq:closed2}: the Friedmann equations and the energy balance are unchanged, and the pairs track $\rhoMH$
instantaneously, with $-\dot\rho_\Lambda=[(3-2\epsilon)\bar H+\Gde]\rhoMH$. This is the cosmic rate equation in the detailed-balance regime $\GaM\gg H$. Without decays
the system has the exact solution $\epsilon=\chi m^2/\mpl^2$ found in Part~I in the detailed-balance regime.

Two points concern the physical interpretation of the closed sets.

\emph{(i) The origin of a finite relaxation rate.} The fast coherent dynamics of Part~II is reversible. It reproduces the cosmic rate
equation only in the regime $\GaM\gg H$, in which the pairs follow $\rhoMH$ without lag. In it, the real pairs never relax towards
$\rhoMH$: they evolve as matter according to Eq.~\eqref{eq:real}. The full closed set \eqref{eq:closed1} contains more than this. Its finite
relaxation rate $\GaM$ describes an irreversible approach of $\rho_M$ to $\rhoMH$, with a lag of order $1/\GaM$. This matters whenever $\GaM$ is
comparable to $H$, as at the end of inflation and during reheating. In the field theory such irreversibility arises only from the coupling
of the pairs to fluctuations of the spacetime vacuum with a continuous spectrum, i.e.\ from the Markov limit of the kinetic description of
section~\ref{sec:kinetic}. The value of $\GaM$ is fixed by the spectral function of these fluctuations at the frequency $2m$, which a
microscopic theory of the spacetime vacuum must provide. The arguments of section~\ref{sec:whym} make the scaling $\GaM\propto m$ plausible but do
not fix its coefficient.

\emph{(ii) The nature of the density $\rhoMH$.} The density $\rhoMH=2\chi m^2H^2$ obtained from the adiabatic occupation is the dressing of the
spacetime vacuum. It follows the expansion instantaneously and would disappear if the expansion stopped. Real pairs left behind by a
smooth expansion are, by contrast, suppressed as $e^{-2\pi m/H}$ (section~\ref{sec:dressing}). In the scenario of
Refs.~\cite{Xue2023,Xue2023b,Xue2024} the pairs of density $\rhoMH$ gravitate as matter and decay into radiation, i.e.\ they behave as a
stationary population of real pairs. Within the present framework, maintaining such a population requires production that is not
exponentially suppressed, through one of three channels:
\begin{itemize}[leftmargin=1.5em,itemsep=2pt]
  \item fluctuations of the spacetime vacuum that absorb more readily than they emit, $S_{\mathcal O}(+2m)>S_{\mathcal O}(-2m)$, which sustain the stationary
  occupation $f^*$ of section~\ref{sec:kinetic};
  \item a drive with spectral weight at $2m$ whose phase is not locked to the pairs (section~\ref{sec:resonance});
  \item dissipation through decays, which converts the dressing into radiation at the rate $\Gde\rhoMH$ (section~\ref{sec:averaged}).
\end{itemize}
Both points lead to the same open question: the spectrum of the fluctuations of the spacetime vacuum at twice the pair mass.

This also marks the limit of the fast-oscillating description of Part~II. There the time dependence of the spacetime vacuum relating to
the cosmological term is represented by a single harmonic oscillation at frequency $2m$ with a slowly varying amplitude,
$\hf{}=\mathrm{Re}[\tilde h\,e^{-2imt}]$ and $\rho_\Lambda^{\rm fast}=6\mpl^2\bar H\hf{}$ [Eqs.~\eqref{eq:hf} and \eqref{eq:fastF}], sourced by the pressure of
the pairs. This representation is adequate in the detailed-balance regime $\GaM\gg H$. There the pairs track $\rhoMH$, the oscillation is
phase-locked to them, and averaging over the fast time gives the closed set \eqref{eq:closed2}.

It is not adequate in general, and in particular not during reheating. In that epoch a plethora of real massive pairs $\rho^{\rm r}$ is
produced, the pairs decay into radiation, and $\GaM\gg H$ is no longer true. The spacetime vacuum then exchanges energy with matter
irreversibly. One should therefore consider more complex time-oscillating behaviours of the spacetime vacuum relating to the cosmological
term than the harmonic one of Eqs.~\eqref{eq:hf}--\eqref{eq:fastF}:
\begin{itemize}[leftmargin=1.5em,itemsep=2pt]
  \item a spectrum of frequencies rather than a single line at $2m$;
  \item phases that are not locked to the pair coherence;
  \item amplitudes that vary on the time scale of reheating itself;
  \item oscillations driven by the populations of real pairs and of their decay products, as well as by the pair coherence.
\end{itemize}
Such behaviour would supply, within a single framework, the continuous spectrum needed for a finite relaxation rate $\GaM$ (point~(i)) and
the production, free of exponential suppression, needed to sustain real massive pairs (point~(ii)). Its study in a theoretical framework of an explicit field theory, together with
the numerical solution of the closed set \eqref{eq:closed1} through the reheating epoch, is left to future work.

\appendix

\section{Kinetic equation and the fluctuation operator}\label{app:kinetic}

\paragraph{Matrix elements.} In a comoving box of volume $V$, the interaction Hamiltonian $H_{\rm int}=\int\dd^3x\,\Ocal_S|\chi|^2$, with
$\Ocal_S$ homogeneous on the scale of the pair wavelength, contains
\begin{equation*}
  \Ocal_S(t)\sum_\kk\big[e^{2i\theta_k}a^\dagger_\kk b^\dagger_{-\kk}+e^{-2i\theta_k}a_\kk b_{-\kk}\big]/2\Omega_k .
\end{equation*}
Between Fock states with $n$ particles in $\kk$ and $\bar n$ antiparticles in $-\kk$, it raises both numbers by one with
amplitude $\Ocal_Se^{2i\theta_k}\sqrt{(n+1)(\bar n+1)}/2\Omega_k$, and lowers both by one with amplitude $\Ocal_Se^{-2i\theta_k}\sqrt{n\bar n}/2\Omega_k$.

\paragraph{Rates.} The slowly varying classical part $\Delta(t)$ has no Fourier component at $2\Omega_k$ and gives only the adiabatic
occupation. For the fluctuating part, the first-order creation amplitude squared and averaged over the state of the spacetime vacuum is
\begin{align*}
  \langle|A_\uparrow|^2\rangle&=\frac{(n+1)(\bar n+1)}{4\Omega_k^2}\int_0^t\!\!\dd t'\!\int_0^t\!\!\dd t''\,e^{2i\Omega_k(t'-t'')}
  \langle\delta\Ocal_S(t'')\delta\Ocal_S(t')\rangle\\
  &\to\;t\,(n+1)(\bar n+1)\,\frac{S_{\mathcal O}(-2\Omega_k)}{4\Omega_k^2},
\end{align*}
where the spacetime-vacuum operator of the conjugate amplitude stands on the left; annihilation follows in the same way. As a check of
the sign, for an oscillator mode of the spacetime vacuum of frequency $W$ and occupation $\bar n_W$ one has
$\langle\delta\Ocal(\tau)\delta\Ocal(0)\rangle\propto(\bar n_W+1)e^{-iW\tau}+\bar n_We^{iW\tau}$. Then $S_{\mathcal O}(+W)$ measures absorption by the spacetime vacuum and
$S_{\mathcal O}(-W)$ emission, and creating a pair requires the spacetime vacuum to supply $2\Omega_k$. We verified this assignment by computing the
first-order transition probability explicitly. A spacetime vacuum in thermal equilibrium at a temperature $T_{\mathcal S}$ satisfies the Kubo--Martin--Schwinger (KMS)
condition~\cite{Kubo1957,MartinSchwinger1959}, $S_{\mathcal O}(-\omega)=e^{-\omega/T_{\mathcal S}}S_{\mathcal O}(\omega)$; it gives a Bose--Einstein stationary occupation with zero pair chemical potential, $\mu_F+\mu_{\bar F}=0$, as in Ref.~\cite{Xue2023b}.

\paragraph{Linearisation and $\GaM$.} The stationary solution of \eqref{eq:Boltz} satisfies $\Gamma^\uparrow(1+f^*)^2=\Gamma^\downarrow f^{*2}$. Writing
$f_{F,\bar F}=f^*+\delta f_{F,\bar F}$ and keeping first-order terms,
$\delta\dot f_F=\delta\dot f_{\bar F}=-\tfrac12\gamma_k(\delta f_F+\delta f_{\bar F})$, with $\gamma_k=2\Gamma^\uparrow_k(1+f^*_k)/f^*_k=2\Gamma^\downarrow_kf^*_k/(1+f^*_k)$. Weighting
with $\Omega_k$ and integrating gives Eq.~\eqref{eq:GaM}, with $\GaM$ the energy-weighted average of $\gamma_k$ (equal to $\gamma$ if $\gamma_k$ is constant
over the shell). For dilute pairs, $f^*=(\Gamma^\uparrow/\Gamma^\downarrow)^{1/2}$ and $\GaM=2(\Gamma^\uparrow\Gamma^\downarrow)^{1/2}$.

\paragraph{The coupling operator in the two cases.} The operator $\Ocal_S=\Delta(t)+\delta\Ocal_S$ in \eqref{eq:La} collects the coupling of $|\chi|^2$ to
the gravitational sector at lowest dimension. In this article it takes two forms.
\begin{itemize}[leftmargin=1.5em,itemsep=2pt]
\item \emph{Adiabatic case (Part~I).} The classical part is $\Ocal_S=\Delta=\xi R-\tfrac94H^2-\tfrac32\dot H$. It contains the curvature coupling $\xi R$
and the terms generated by the rescaling $\chi=a^{3/2}\Phi$. It varies on the Hubble time and gives the adiabatic occupation and $\rhoMH$
(section~\ref{sec:rhoH}). The fluctuations $\delta\Ocal_S$ of the spacetime vacuum around this classical value, with a continuous spectrum,
give the kinetic equation and the relaxation rate $\GaM$ (section~\ref{sec:kinetic}).
\item \emph{Fast-oscillating case (Part~II).} The fluctuation is the fast coherent oscillation of the Hubble function,
$\delta\Ocal_S=(6\xi-\tfrac32)\dot h_f^{\rm fast}$, a deterministic function with a spectral line at $\pm2m$ and a symmetric spectrum
(section~\ref{sec:resonance}).
\end{itemize}
Only the two-point function of $\delta\Ocal_S$ enters at the order considered. The kinetic derivation assumes stationarity, a correlation
time short compared with $\GaM^{-1}$ (the Markov limit), and $\langle\delta\Ocal_S\rangle=0$. A classical source gives $\Gamma^\uparrow=\Gamma^\downarrow$ and no stationary
state. A finite $\rhoMH$ in the kinetic description therefore requires the quantum asymmetry $S_{\mathcal O}(+2m)>S_{\mathcal O}(-2m)$ of the
fluctuations of the spacetime vacuum.

\paragraph{Spatial correlations of the spacetime vacuum.} If $\delta\Ocal_S$ carries momentum, a particle and an antiparticle can be
created with different momenta $\kk$ and $\kk'$, i.e.\ with total momentum $\kk+\kk'$. The kernel of the kinetic equation is then
\begin{equation*}
  W^{\uparrow,\downarrow}_{\kk\kk'}=\frac{S_{\mathcal O}(\mp(\Omega_k+\Omega_{k'}),\kk+\kk')}{4\Omega_k\Omega_{k'}},\qquad
  S_{\mathcal O}(\omega,\mathbf q)=\int\!\dd\tau\,\dd^3r\,e^{i\omega\tau-i\mathbf q\cdot\mathbf r}\langle\delta\Ocal_S(\tau,\mathbf r)\delta\Ocal_S(0,\mathbf 0)\rangle ,
\end{equation*}
where $S_{\mathcal O}(\omega,\mathbf q)$ is dimensionless. In the local
limit, in which the correlation length is short compared with the pair wavelength, every particle can annihilate with every
antiparticle. The annihilation term of the kinetic equation becomes $\langle\sigma v\rangle n_Fn_{\bar F}$, with $\langle\sigma v\rangle=S_{\mathcal O}(2m,\mathbf0)/4m^2$, and the
linearised relaxation rate is $\GaM=2\langle\sigma v\rangle\nMH$.

Reproducing the estimate $\GaM=\chi m\epsilon/4\pi$ with $\nMH=\chi mH^2$ then requires
\begin{equation}
  \langle\sigma v\rangle=\frac{\GaM}{2\nMH}=\frac{\epsilon}{8\pi H^2},
  \label{eq:sigmav}
\end{equation}
an annihilation cross section of the order of the horizon area, independent of $m$ and $\chi$. To see what this means, compare it
with the Compton area $m^{-2}$ that a perturbative local coupling would give. Take the reheating epoch, with
$H\approx10^{-5}$--$10^{-9}\,\mpl$, and $m=27.7\,\mpl$~\cite{Xue2025}. The Hubble radius then contains $m/H\approx3\times10^6$--$3\times10^{10}$ Compton lengths,
and \eqref{eq:sigmav} gives $\langle\sigma v\rangle m^2=(\epsilon/8\pi)(m/H)^2\approx5\times10^{11}$--$5\times10^{19}$ for $\epsilon=3/2$.

A local coupling with $\delta\Ocal_S\lesssim m^2$ and correlation time $\sim m^{-1}$ gives $S_{\mathcal O}(2m,\mathbf0)\sim(m\ell)^3$, where $\ell$ is the correlation length of the
fluctuations. Matching \eqref{eq:sigmav} requires
$m\ell\sim[4\langle\sigma v\rangle m^2]^{1/3}\approx10^4$--$6\times10^6$, i.e.\ $\ell\approx(4\times10^{-3}$--$2\times10^{-4})\,H^{-1}$ for $H=10^{-5}$--$10^{-9}\,\mpl$:
\begin{center}\small
\begin{tabular}{@{}ccccc@{}}
\toprule
$H/\mpl$ & $m/H$ (Compton lengths per $H^{-1}$) & $\langle\sigma v\rangle m^2$ & $m\ell$ & $\ell H$\\
\midrule
$10^{-5}$ & $2.8\times10^{6}$ & $4.6\times10^{11}$ & $1.2\times10^{4}$ & $4.4\times10^{-3}$\\
$10^{-7}$ & $2.8\times10^{8}$ & $4.6\times10^{15}$ & $2.6\times10^{5}$ & $9.5\times10^{-4}$\\
$10^{-9}$ & $2.8\times10^{10}$ & $4.6\times10^{19}$ & $5.7\times10^{6}$ & $2.1\times10^{-4}$\\
\bottomrule
\end{tabular}
\end{center}
The fluctuations of the spacetime vacuum must therefore be strongly coupled and correlated over $10^4$--$10^7$ Compton lengths. This
is far beyond the range of a local perturbative coupling, yet still well inside the Hubble radius, at $10^{-2}$--$10^{-4}$ of it. The
correlation length decreases relative to the horizon as $H$ decreases, $\ell H\propto(H/m)^{1/3}$. The required correlations are thus
mesoscopic: intermediate between the Compton scale of the pairs and the horizon scale. This is consistent with the holographic
interpretation of the pair plasma in Refs.~\cite{Xue2023b,Xue2024}, in which the pairs are produced and oscillate near the horizon,
but not with a weak local interaction.

\section{Two-timescale expansion}\label{app:twotime}

\paragraph{Real form and structure.} With $c_k=u_k+iv_k$, Eq.~\eqref{eq:kin} becomes
$\dot f_k=(\dot\Omega_k/\Omega_k)u_k$, $\dot u_k=2\Omega_kv_k+(\dot\Omega_k/\Omega_k)f_k+\dot\Omega_k/2\Omega_k$ and $\dot v_k=-2\Omega_ku_k$: a linear inhomogeneous system for
three real variables. The decomposition \eqref{eq:split} is unique once the adiabatic part is required to be local in the
background, $f^{\rm r}$ to be slow, and the oscillating parts to have zero fast average. Operationally, $f^{\rm osc}=f-\langle f\rangle$,
$f^{\rm r}=\langle f\rangle-f^{\rm ad}$ and $c^{\rm osc}=c-c^{\rm ad}$.

\paragraph{Adiabatic invariance to all orders.} Attach a factor $\delta$ to each time derivative of the background and expand
$u=\sum\delta^nu_n$, $v=\sum\delta^nv_n$, $f=f^{\rm r}+\sum_{n\ge1}\delta^nf_n$. Order by order,
\begin{equation*}
  v_n=\frac{1}{2\Omega}\Big[\dot u_{n-1}-\frac{\dot\Omega}{\Omega}f_{n-1}-\frac{\dot\Omega}{2\Omega}\delta_{n1}\Big],\qquad
  u_n=-\frac{\dot v_{n-1}}{2\Omega},\qquad \dot f_n=\frac{\dot\Omega}{\Omega}u_n ,
\end{equation*}
with $u_0=v_0=0$. The odd orders vanish. At even orders, the Euler--Lagrange operator of $\dot f_n$ with respect to $\Omega$ vanishes
identically, so $\dot f_n$ is the total time derivative of a local function. We verified this symbolically through sixth order and
constructed the antiderivatives:
\begin{equation*}
  f_2=(1+2f^{\rm r})\Big(\frac{\dot\Omega}{4\Omega^2}\Big)^2,\qquad
  f_4=(1+2f^{\rm r})\,\frac{4\Omega^2\ddot\Omega^2-8\Omega^2\dot\Omega\dddot\Omega+40\Omega\dot\Omega^2\ddot\Omega-45\dot\Omega^4}{256\,\Omega^8}.
\end{equation*}
Hence $\langle f_k\rangle=f^{\rm r}_k+(1+2f^{\rm r}_k)\mathcal A_k(t)$. Here
\begin{equation*}
  \mathcal A_k(t)\equiv\sum_{j\ge1}\frac{f_{2j}}{1+2f^{\rm r}_k}=\Big(\frac{\dot\Omega_k}{4\Omega_k^2}\Big)^2+\mathcal A_k^{(4)}+\mathcal A_k^{(6)}+\cdots
\end{equation*}
is the adiabatic occupation per unit Bose factor $1+2f^{\rm r}_k$: the sum of the even-order terms of the expansion, each a local function
of $\Omega_k$ and its time derivatives at time $t$, with $\mathcal A_k^{(4)}=f_4/(1+2f^{\rm r})$ given above. Thus $f_k^{\rm ad}=(1+2f^{\rm r}_k)\mathcal A_k$, whose
leading term is Eq.~\eqref{eq:fad}. Since $\mathcal A_k$ is local, $(1+2\langle f_k\rangle)/(1+2\mathcal A_k)$ is a first integral. The number of
quanta changes only through effects outside the adiabatic series: non-perturbative creation, resonance with $\hf{}$, dissipation,
or irreversible exchange with the spacetime vacuum.

\paragraph{Oscillating part.} The fast homogeneous solutions are $c^{\rm osc}_k=|c^0_k|e^{-2i\theta_k+i\varphi_k}[1+\Ocal(\zeta_k)]$. Inserting them into
$\dot f_k=(\dot\Omega_k/\Omega_k)\mathrm{Re}\,c_k$ and integrating by parts with $\dd\theta_k/\dd t=\Omega_k$ gives
$f^{\rm osc}_k=(\dot\Omega_k/2\Omega_k^2)|c^0_k|\sin(2\theta_k-\varphi_k)$. Because $|c_k|^2-f_k(1+f_k)$ is conserved, $|c^{\rm osc}_k|^2\simeq f^{\rm r}_k(1+f^{\rm r}_k)$ for the
pure state.

\paragraph{Averaged pair equation.} Exactly,
\begin{equation*}
  \frac{\dd(a^3\rho_M)}{\dd t}=2g_\Phi\!\int\!\frac{\dd^3k}{(2\pi)^3}\big(\dot\Omega_kf_k+\Omega_k\dot f_k\big)
  =2g_\Phi\!\int\!\frac{\dd^3k}{(2\pi)^3}\,\dot\Omega_k\big(f_k+\mathrm{Re}\,c_k\big).
\end{equation*}
For the adiabatic terms, $\Omega_k\dot f^{\rm ad}_k=\dot\Omega_k\mathrm{Re}\,c^{\rm ad}_k$, so that
$\dot\Omega_k(f^{\rm ad}_k+\mathrm{Re}\,c^{\rm ad}_k)=\dd(\Omega_kf^{\rm ad}_k)/\dd t$. The real-pair term, with $\dot\Omega_k=-\bar Hp^2/\Omega_k$, gives $-3\bar Ha^3w^{\rm r}\rho^{\rm r}$. The
oscillating terms average to zero, and cross terms between the fast part of $\dot\Omega$ and $f^{\rm osc}$ are $\Ocal(\zeta_k^2)$. With decays, the
slaved coherence $c_k\simeq(\dot\Omega_k/2\Omega_k)(1+2f_k)/(2i\Omega_k+\tfrac12\Gde)$ has real part $\Gde\dot\Omega_k(1+2f_k)/16\Omega_k^3$. This gives the replenishment
$\Gde f^{\rm ad}_k$ quoted in section~\ref{sec:averaged}.

\paragraph{Slaved solution of the rate equation.} With $\rho_M=\rhoMH+\delta$, the rate equation \eqref{eq:rate} becomes $\dot\delta+\Gtot{}\delta=-S$.
Iterating $\delta=-(S+\dot\delta)/\Gtot{}$ gives $\delta_{\rm s}=-S/\Gtot{}+\Gtot^{-1}\dd_t(S/\Gtot{})+\cdots$, and the free part is $\delta_{\rm h}=\delta_0\exp\!\big(-\!\int^t\Gtot\,\dd t'\big)$. Hence
$\GaM(\rhoMH-\rho_M)=S(\GaM/\Gtot{})[1+\Ocal(H/\Gtot{})]-\GaM\delta_{\rm h}$. Expanding $\GaM/\Gtot{}=1-(3H+\Gde)/\GaM+\Ocal(\GaM^{-2})$ gives the correction quoted in
section~\ref{sec:closed2}.

\section{Symbolic checks}\label{app:checks}

The following results were verified with computer algebra: the Christoffel symbols and the Einstein tensor \eqref{eq:geometry};
the rescaled action \eqref{eq:Schi}; the kinetic equations \eqref{eq:kin} and the expectation values behind \eqref{eq:rhop}; the fast coupling
$(6\xi-\tfrac32)\dot h_f^{\rm fast}$ and the resonant coefficient \eqref{eq:g}; the Bogoliubov solution \eqref{eq:nk} and the conservation of the
particle-minus-antiparticle number; the density and pressure \eqref{eq:rhop_osc}; the adiabatic expansion through sixth order and
the antiderivatives $f_2$ and $f_4$ (appendix~\ref{app:twotime}); and the exact solution \eqref{eq:exact}. The numerical results quoted
in sections~\ref{sec:resonance}, \ref{sec:dressing}, \ref{sec:averaged} and \ref{sec:closed2} were obtained by direct integration of the kinetic
equations \eqref{eq:kin}, the pump equations \eqref{eq:pump} and the rate equation \eqref{eq:rate}.

\bibliographystyle{JHEP}
\bibliography{refs}
\end{document}